\documentclass[pra,reprint,doublecolumn,superscriptaddress, nobalancelastpage,dvipsnames]{revtex4-2}
\usepackage[dvipsnames]{xcolor}

\usepackage{tikz}
\usepackage{quantikz}
\usetikzlibrary{shapes.misc, positioning, backgrounds}

\usepackage[sort&compress]{natbib}
\usepackage{verbatim}
\usepackage{float}
\usepackage{sidecap}
\sidecaptionvpos{figure}{c}
\usepackage{lipsum}
\usepackage{placeins}
\usepackage{capt-of}
\usepackage{setspace}
\usepackage{dcolumn}
\usepackage{bm}
\usepackage{amssymb}
\usepackage{amsmath}
\usepackage{mathtools}
\usepackage{booktabs}
\usepackage{physics}
\usepackage{multirow}
\usepackage[a4paper]{geometry}
\newgeometry{top=1.25cm,right=1cm,left=1.25cm,bottom=1.25cm}
\usepackage[colorlinks=true, urlcolor=blue, pdfborder={0 0 0}]{hyperref}
\hypersetup{linkcolor=blue,citecolor=BrickRed}

\begin{document}

\title{\Large{\textbf{Calibrating the role of entanglement in variational quantum circuits}}}

\author{Azar C. Nakhl} \email{a.nakhl@student.unimelb.edu.au} \affiliation{School of Physics, The University of Melbourne, Parkville, 3010, VIC, Australia}
\author{Thomas Quella} \email{Thomas.Quella@unimelb.edu.au} \affiliation{School of Mathematics and Statistics, The University of Melbourne, Parkville, 3010, VIC, Australia}
\author{Muhammad Usman} \email{musman@unimelb.edu.au}  \affiliation{School of Physics, The University of Melbourne, Parkville, 3010, VIC, Australia}
\affiliation{Data61, CSIRO, Clayton, 3168, VIC, Australia}

\maketitle%

\onecolumngrid%

\noindent
\textcolor{black}{\normalsize{\textbf{Entanglement is a key property of quantum computing that separates it from its classical counterpart, however, its exact role in the performance of quantum algorithms, especially variational quantum algorithms, is not well understood. In this work, we utilise tensor network methods to systematically probe the role of entanglement in the working of two variational quantum algorithms, the Quantum Approximate Optimisation Algorithm (QAOA) and Quantum Neural Networks (QNNs), on prototypical problems under controlled entanglement environments. We find that for the MAX-CUT problem solved using QAOA, the fidelity as a function of entanglement is highly dependent on the number of layers, layout of edges in the graph, and edge density, generally exhibiting that a high number of layers indicates a higher resilience to truncation of entanglement. This is in contrast to previous studies based on no more than four QAOA layers which show that the fidelity of QAOA follows a scaling law for circuits. Contrarily, in the case of QNNs, trained circuits with high test accuracies are underpinned by higher entanglement, with any enforced limitation in entanglement resulting in a sharp decline in test accuracy. This is corroborated by the entanglement entropy of these circuits which is consistently high suggesting that, unlike QAOA, QNNs may require quantum devices capable of generating highly entangled states. Overall our work provides a deeper understanding of the role of entanglement in the working of variational quantum algorithms which may help to implement these algorithms on NISQ-era quantum hardware in a way that maximises their accuracies.
  }}}
\\ \\ \\
\twocolumngrid%
\section{Introduction}

The development and benchmarking of quantum algorithms have attracted significant attention in recent years because their in-principle capability to solve classically intractable problems\cite{nielsen_quantum_2012} when coupled with advances in quantum hardware is anticipated to provide a quantum advantage in a range of real-world applications\cite{preskill_quantum_2018}. Among a variety of quantum algorithm classes that are being developed, one particular class known as Variational Quantum Algorithms (VQA) has been the subject of intense research due to the possibility of combining the manipulation of quantum systems with classical optimisation techniques\cite{peruzzo_variational_2014}, allowing one to take advantage of the high dimensionality of the state space of quantum systems along with sophisticated optimisation algorithms. In particular, these quantum algorithms with classically optimised variational parameters have found applications in chemistry\cite{kandala_hardware-efficient_2017,fedorov_vqe_2022,jones_chemistry_2022}, finance\cite{egger_quantum_2020,orus_quantum_2019}, operations research\cite{bentley_quantum_2022,harwood_formulating_2021,azad_solving_2023,vikstal_applying_2020} and more recently in quantum machine learning models\cite{cerezo_challenges_2022,farhi_classification_2018,romero_quantum_2017,west_benchmarking_2023,west_towards_2023,daskin_simple_2018,bisarya_breast_2020,sagingalieva_hybrid_2023,li_model_2014}.

However, unlike conventional quantum algorithms, such as those for dictionary search\cite{grover_fast_1996}, or semi-prime number factorisation\cite{shor_polynomial-time_1997}, VQAs rely on heuristic approaches. Importantly this means that even without quantum hardware noise there is no guarantee of these algorithms' performance on any given instance. As a result, there is a significant amount of research on the performance analysis of these algorithms\cite{farhi_quantum_2020,stilck_franca_limitations_2021,gonthier_measurements_2022,wang_noise-induced_2021,PRXQuantum.4.010309,PhysRevB.106.214429,PRXQuantum.1.020319}.
Entanglement is one of the key metrics that may be used to analyse the performance of VQAs, but as of yet there has only been limited investigation of the role of entanglement with regard to these algorithms\cite{dupont_calibrating_2022,dupont_entanglement_2022,ballarin_entanglement_2023,PhysRevA.104.062426,chen2022much}. 
Our work aims to fill this knowledge gap by providing a systematic and comprehensive calibration of the role of entanglement in the functionality of two widely used variational quantum algorithms: (1) the Quantum Approximate Optimisation Algorithm (QAOA)\cite{farhi_quantum_2014} used for combinatorial optimisation tasks and; (2) Quantum Neural Networks (QNNs)\cite{farhi_classification_2018} used in a variety of machine learning applications including in image classification\cite{farhi_classification_2018,west_benchmarking_2023,west_towards_2023}, drug response prognosis\cite{sagingalieva_hybrid_2023} and breast cancer prediction\cite{li_model_2014,daskin_simple_2018,bisarya_breast_2020}, among others. This work will use Matrix Product State\cite{vidal_efficient_2003,schollwock_density-matrix_2011} (MPS) simulations to analyse how these two algorithms perform when their entanglement is restricted. We will also track the entanglement entropy of these algorithms as the system size and depth of the quantum circuit increase.
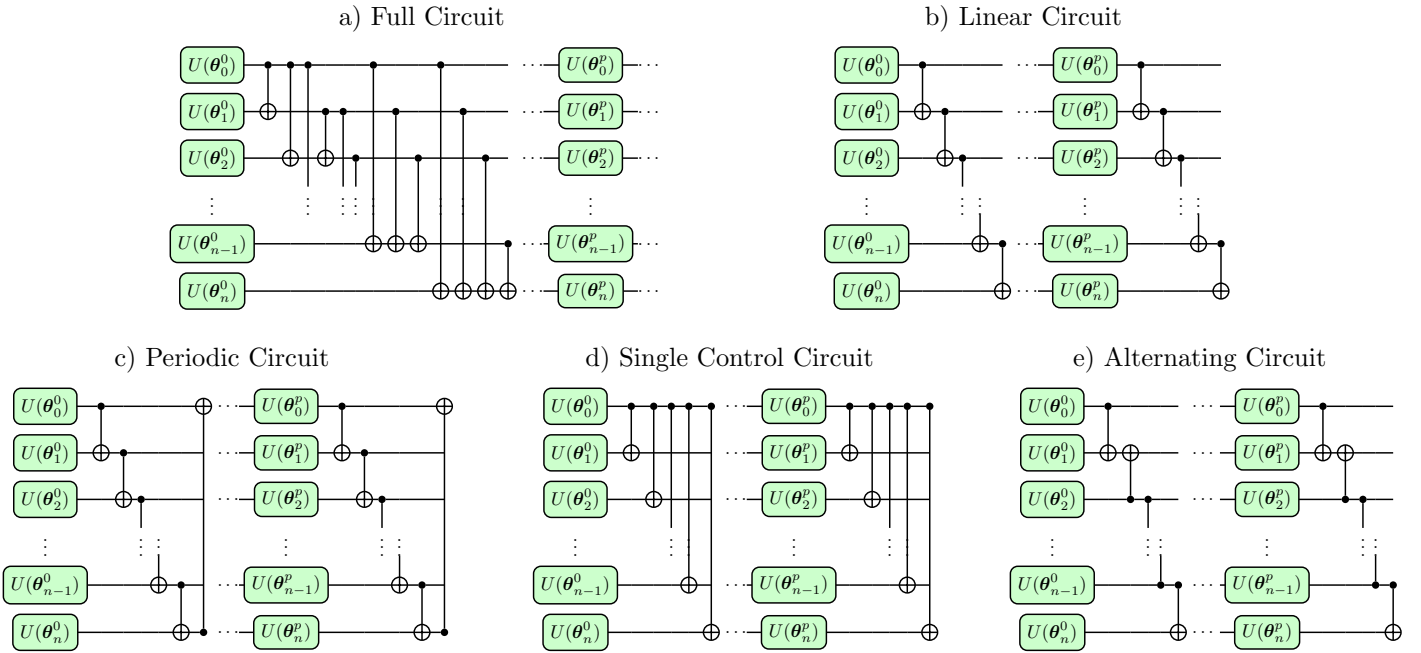
\begin{figure*}
    \begin{tikzpicture}
        \node at (0,0) {
            \includegraphics[scale=0.7]{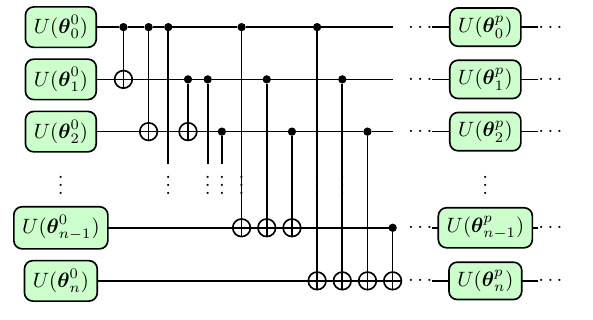}
        };
        \node at (8,0) {
            \includegraphics[scale=0.7]{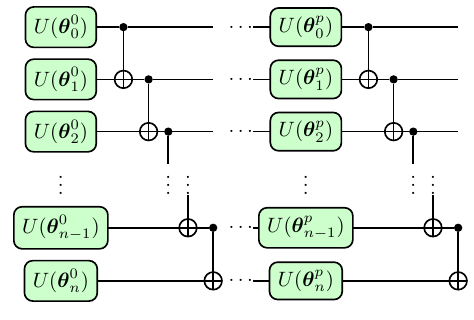}
        };
        \node at (0,2.2) {
           a) Full Circuit 
        };
        \node at (8,2.2) {
           b) Linear Circuit 
        };
    \end{tikzpicture}
    \begin{tikzpicture}
        \node at (0,0) {
            \includegraphics[scale=0.7]{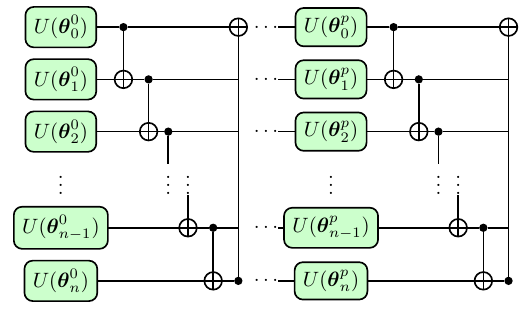}
        };
        \node at (6.75,0) {
            \includegraphics[scale=0.7]{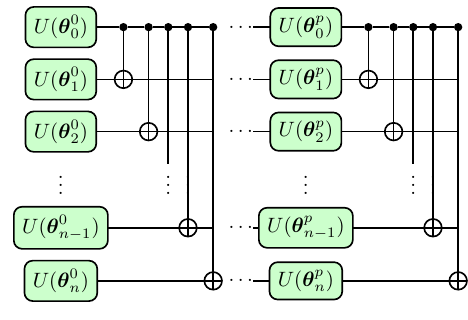}
        };
        \node at (13,0) {
            \includegraphics[scale=0.7]{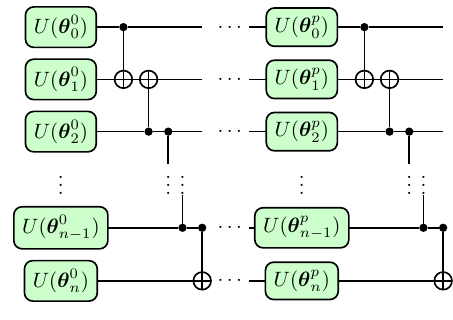}
        };
        \node at (0,2.2) {
           c) Periodic Circuit 
        };
        \node at (6.75,2.2) {
            d) Single Control Circuit
        };
        \node at (13,2.2) {
            e) Alternating Circuit
        };
    \end{tikzpicture}
    \caption{The various QNN circuits considered in this work with labels above. Each block may be repeated $p$ times with new parameter vectors $\boldsymbol{\theta}_i^j$ each taking three real values. For an $n$ qubit system, each layer has $n$ $U(\mathbf{\theta})$ arbitrary one-qubit gates as defined in Ref. \cite{nielsen_quantum_2012}. The number of entangling gates per layer is a) $n(n-1)/2$, b) $n$, c) $n+1$, d)-e) $n$. All circuits have the first and last layers shown, except for the full circuit for brevity. The different circuits represent a different entanglement structure, except for the Linear and Full circuits which can be shown to be identical as per Ref. \cite{ballarin_entanglement_2023}.}
    \label{fig:qml-circs}
\end{figure*}

 For QAOA, in the context of the paradigmatic combinatorial optimisation problem MAX-CUT, it has been shown that there is a volume law entanglement barrier between the initial and final state of the algorithm\cite{dupont_entanglement_2022}, suggesting that tensor network simulation methods\cite{vidal_efficient_2003,schollwock_density-matrix_2011}, whose memory and computational resource requirements scale exponentially with the entanglement of the quantum system are not able to simulate QAOA efficiently. This is corroborated by Ref. \cite{dupont_calibrating_2022} which shows that for complete graphs with random edge weights and 3-regular graphs with random edges, the fidelity of QAOA with few number of layers follows a scaling law with respect to the entanglement per qubit such that the fidelity is the same regardless of the size of the system. It remains open as to whether such a scaling law holds for circuits with greater than four QAOA layers or different classes of graphs, such as grid graphs or $k$-regular graphs (where $k\neq3$), a question that we will address in this work. Similar studies of the fidelity with respect to entanglement per qubit have also been conducted for Grover's Algorithm, the Quantum Fourier Transform and the quantum counting algorithm\cite{niedermeier_tensor-network_2023}.

A second class of VQAs that has attracted significant attention in recent years are Quantum Neural Networks (QNNs) used to perform machine learning tasks. In general, Quantum Machine Learning (QML) models can range from quantum subroutines of an overall classical process\cite{farhi_classification_2018,romero_quantum_2017,west_towards_2023,west_benchmarking_2023,daskin_simple_2018,bisarya_breast_2020,li_model_2014,sagingalieva_hybrid_2023} to completely quantum analogues of established machine learning models\cite{havlicek_supervised_2019,srikumar_kernel-based_2023}. Contrary to other VQAs however there has only been limited investigation as to how entanglement grows within these QML models. For example, in the case of parameter optimisation-based methods such as QNNs, it is recognised that entanglement-induced barren plateaus will arise for models that satisfy volume-law growth in their entanglement entropy\cite{ortiz_marrero_entanglement-induced_2021}. Additionally, there has also been investigation of the entanglement entropy of QNN architectures with random parameters for up to 50 qubits\cite{ballarin_entanglement_2023}. However, we are not aware of any previous work which has studied the entanglement of trained QNNs. This will be the second focus of our work.

Tensor Networks provide an efficient method to quantitatively assess and control the entanglement of a quantum state. In particular, MPS have found significant application in the simulation of quantum algorithms on classical computers in general\cite{dang_distributed_2017,banuls_simulation_2006}, and specifically for the study of entanglement in quantum algorithms \cite{dupont_calibrating_2022,niedermeier_tensor-network_2023,ballarin_entanglement_2023}. MPS are characterised by one freely adjustable parameter, their bond dimension, which translates into an upper bound on the entanglement entropy of the quantum system. Moreover, the entanglement entropy can be efficiently determined from an MPS\cite{orus_practical_2014}. 

The bond dimension $\chi$ of an MPS is the primary driver of memory utilisation, scaling as $O(N\chi^2)$ for a system with $N$ sites. As a result, states of high entanglement are difficult to represent, whereas systems with relatively low entanglement may be represented in an efficient manner.
This may be compared to state-vector simulations which generally have a memory requirement of $O(2^N)$\cite{nielsen_quantum_2012}. Unlike state-vector simulation methods, quantum states represented as MPS may be approximated in such a way that their bond dimension is reduced systematically, providing a way to control the entanglement of the state at the cost of fidelity. This, along with the ability to perform local operations on the state locally in memory, are the key benefits of using MPS to study quantum systems and quantum algorithms.

In this work, we study the properties of QAOA when applied to the solution of the MAX-CUT problem on 3-regular, 4-regular, complete and grid graphs for a varying number of QAOA layers. We demonstrate that the entanglement of the resulting QAOA circuits is highly dependent on both the type of graph and the number of QAOA layers. As a result, we find that previously proposed scaling laws for 3-regular and complete graphs\cite{dupont_calibrating_2022}
do not hold true for larger-depth QAOA circuits or graphs with a structured edge layout such as the grid graph. This observation has implications for the use of QAOA to solve tasks, such as the vehicle routing problem\cite{azad_solving_2023} which has an underlying graph that is low density and regular, and some spin-glass models\cite{SFEdwards_1975} where the underlying graph is a grid graph. We find that the entanglement per qubit required to achieve high fidelity for such types of underlying graphs is low. This is contrasted with tasks such as aircraft tail assignment\cite{vikstal_applying_2020} which have underlying graphs which are of high density, where we conclude that the entanglement per qubit required to achieve high fidelity is relatively high. 

Importantly, our findings demonstrate that the entanglement scaling of QAOA on average is not indicative of its scaling for specific tasks and furthermore that the entanglement at low depth, where one is unlikely to find an appropriate solution is fundamentally different to that at high depth, where our work suggests that the entanglement per qubit required to achieve high fidelity is significantly reduced.

We then consider QNNs which are a type of VQA that have a circuit structure which is substantially different from that of QAOA. We show that various QNN ansätze, each corresponding to a different entangling gate layout as per FIG. \ref{fig:qml-circs}, have an entanglement entropy that depends on the dataset that is considered. In general, we find that the QNNs are highly entangled, with simpler datasets having a relatively lower entropy of entanglement throughout the QNN. We show that the test accuracy scaling as a function of the entanglement per qubit is strongly correlated with the accuracy of the training, with poorly trained circuits being more resilient to a truncation in the entanglement of the circuit. This indicates a high degree of entanglement present in the trained circuits, suggesting that a pathway to demonstrating quantum advantage for QNNs may be available. 

\section{Methods}
In this section, the theory underpinning the MPS representation of quantum systems will be introduced, focusing on how its key characteristic, the bond dimension is related to a quantum state and how it can be used as a metric for entanglement. The bond dimension and control thereof will be central to our analysis of QAOA and QNNs. QAOA will also be introduced, along with the specific problems that will be considered, namely MAX-CUT on varying types of regular, complete and grid graphs. Lastly QNNs, and specifically their application to image classification will be considered, where different types of circuits will be introduced, along with different datasets on which our circuits will be trained. QAOA and QNNs have a wide variety of applications and encapsulate two of the most common circuit architectures for VQAs, where for QAOA the circuit structure is necessarily one that is dependent on the problem instance, whereas for QNNs the circuit structure may be generic or tailored to the quantum device which the algorithm will be executed on.

\subsection{Matrix Product States} \label{sec:mps}
MPS are a tensor network representation of one-dimensional many-body quantum states. We will assume that the total system consists of $N$ individual $d$-level subsystems whose Hilbert space is spanned by orthonormal vectors $|i_s\rangle$ with $i_s=0,\ldots,d-1$ and $s=1,\ldots,N$. For qubits, the relevant unit in quantum computing applications, one has $d=2$. Each many-body quantum state can then be represented in the form,
\begin{equation}
    |\psi\rangle = \sum_{i_1, i_2, \dots, i_N} A^{(1)}_{i_1} A^{(2)}_{i_2} \dots A^{(N)}_{i_N} |i_1 i_2 \dots i_N \rangle,
    \label{eq:mps}
\end{equation}
where the $A^{(s)}_{i_s}$ are suitably sized matrices that encode its entanglement. It is custom to choose these matrices to be square matrices of $s$-independent size $\chi\times\chi$, where $\chi$ is referred to as the bond dimension. Only the matrices at the two ends of the matrix product are $1\times\chi$ and $\chi\times1$ matrices, respectively. This ensures that the matrix product evaluates to a complex number in Eq.\ \eqref{eq:mps}. Alternatively, for fixed $s$ the set of matrices $A^{(s)}_{i_s}$ can be viewed as a rank-3 tensor, with one physical index $i_s$ and two auxiliary indices corresponding to the matrix rows and columns. Using Penrose graphical notation\cite{orus_practical_2014} MPS are then represented as,
\begin{equation}
    \ket{\psi} = \begin{array}{c}
                    \centering\includegraphics[scale=1]{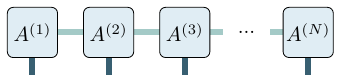}\;,
                 \end{array}
\end{equation}
where $A^{(s)}= \sum_{i = 0,1,\dots,d-1} \ket{i}\otimes A^{(s)}_{i}$ is a rank-3 tensor.

Given an MPS one may determine the entanglement entropy of a bipartition between sites $s$ and $s+1$ by performing high-order singular value decomposition (SVD)\cite{bergqvist_higher-order_2010} on the $A^{(s)}$. Performing a high order SVD on all the $A^{(s)}$ allows for the MPS to be put into a canonical form\cite{perez-garcia_matrix_2007} in which it is a simple process to determine the entanglement entropy of the state at any given bipartition between adjacent sites. The singular values $\Lambda_i^{(s)}$ may be used in the following equation to give the entanglement entropy,
\begin{equation}
    S_s = -\sum_i \abs{\Lambda^{(s)}_i}^2\log(\abs{\Lambda^{(s)}_i}^2) \label{eq:entrop}.
\end{equation}
By the principle of maximum entropy, $S_s$ is maximal when the $(\Lambda^{(s)}_i)^2=\frac{1}{\chi}$ for all $i$. When this is the case one finds $S_s=\log\chi$, which is an upper bound for the entanglement entropy of any bipartition. Hence any limitation in $\chi$ is effectively enforcing a reduction in the entanglement entropy of the bipartition. We will use this to our advantage to produce low entanglement approximations of quantum states at the cost of a reduction in the fidelity of the state. 

One can enforce a reduction in the bond dimension of an MPS via a process known as truncation. As part of this process, the smaller singular values resulting from the higher-order SVD on the $A^{(s)}$ are set to zero. In doing so the bond dimension $\chi$ of the MPS is reduced. A more detailed outline of the process of truncation can be found in Ref. \cite{orus_practical_2014}. The result of this process is an MPS which only approximates the original state. 

For our MPS-based circuit simulations, we utilise the open-source library \texttt{quimb}\cite{gray_quimb_2018} which has in-built quantum circuit simulation capability. Local expectation values are computed analytically at the end of each circuit. Due to performance constraints with the \texttt{quimb} library, the VQA minimisation procedure is performed using a conventional state vector simulator, specifically that packaged within IBM's Qiskit SDK\cite{treinish_qiskitqiskit-metapackage_2023} for QAOA and Xanadu's Pennylane\cite{bergholm_pennylane_2022} for QML. This is equivalent to performing the minimisation using an MPS-based circuit simulator without truncation. This procedural limitation is noted in the results, with the implications of such limitations considered for both VQAs studied. The VQAs are then executed with the optimised angles using an MPS simulator with truncation occurring at various points throughout the circuit as specified in the upcoming subsections. Circuits are truncated by keeping only $2^i$ singular values per internal bond for all $i=1,2,\dots,{N/2}$. As such one has approximations ranging from low entanglement to high entanglement. It is recognised that in general $2^{N/2}$ singular values are required to represent an $N$ qubit system as an MPS exactly. For the truncated circuits, we thus use the following metric to illustrate the entanglement of the truncated circuit,
\begin{equation}
    \text{Entanglement per Qubit} = \frac{2\log_2(\chi)}{N}.
\end{equation}
This metric is the same as that used in Ref. \cite{dupont_calibrating_2022}, rescaled such that it is in the interval $[0,1]$. This is done so that the value measured by this metric is more intuitively understood without the need to consider the underlying MPS representation. To quantify the performance of truncated MPS simulations, we define a quantity that we call the simulation fidelity which is identical to the ``fidelity'' metric used in Ref. \cite{dupont_calibrating_2022} as,
\begin{equation}
    \text{Simulation Fidelity}  = \frac{\text{Cost}(\chi)}{\text{Cost}(2^{N/2})} \label{eq:sim_fid}.
\end{equation}
We call the metric the simulation fidelity to avoid confusion with the conventional definition of fidelity used for quantum states. The $\text{Cost}(\chi)$ is the cost function associated with the VQA evaluated using an MPS circuit simulation which has a maximum bond dimension of $\chi$. This will be explicitly defined for the VQAs where this metric is used in the sections below. A further discussion on the simulation fidelity, particularly as it pertains to this work can be found in Appendix \ref{apn_sim_fid}.
\subsection{Quantum Approximate Optimisation Algorithm}\label{sec:qaoa_method}
QAOA prepares a quantum circuit as a sequence of time-evolution operators of some problem Hamiltonian $H_P$, representing some combinatorial optimisation problem, and a mixer Hamiltonian $H_M$\cite{farhi_quantum_2014}. For an $n$-qubit system, a $p$ layered QAOA takes the following form,
\begin{equation}
    \ket{\psi(\alpha_i, \beta_i)} = \prod_{k=1}^p (e^{-i \alpha_k H_P}e^{-i \beta_k H_M}) \ket{+}^{\otimes n}, \label{eq:qaoa}
\end{equation}
where the product is ordered in the following way, $\prod_{k=1}^p A_k=A_p A_{p-1} \dots A_1$. The initial state is 
defined by $\ket{+}=\frac{1}{\sqrt{2}}(\ket{0}+\ket{1})$, and the mixer Hamiltonian is defined as $H_M=\sum_{i=1}^n X_i$, where $X$ is the Pauli $X$ gate. Note that the problem Hamiltonian $H_P$ is assumed to be diagonal in the computational basis, and is, for the purposes of this work, quadratic at most. As such the state described by Equation (\ref{eq:qaoa}) may be prepared in a straightforward manner using standard one and two-qubit operators. The parameters $\alpha_i$, $\beta_i$ in Equation \eqref{eq:qaoa} may then be optimised using the following cost function,
\begin{equation}
    \text{Cost}(\alpha_i, \beta_i, \chi) = \mel{\psi_\chi(\alpha_i, \beta_i)}{H_P}{\psi_\chi(\alpha_i, \beta_i)},
\end{equation}
where $\ket{\psi_\chi}$ indicates an approximation of the state $\ket{\psi}$ with bond dimension of at most $\chi$ performed using truncation as per Section \ref{sec:mps}.
\begin{figure*}
    \begin{tikzpicture}
        \node at (-1.2,1.2) {a)};
        \node at (0,0) {
            \includegraphics{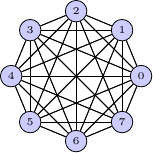}
        };
        \node at (1.7,1.2) {b)};
        \node at (3,0) {
            \includegraphics{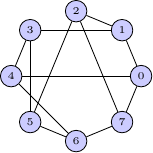}
        };
        \node at (4.7,1.2) {c)};
        \node at (6,0) {
            \includegraphics{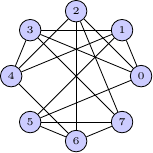}
        };
        \node at (7.8,1.2) {d)};
        \node at (10.1,1.2) {e)};
        \node at (13.4,1.2) {f)};
        \node at (12.5,0) {
            \includegraphics{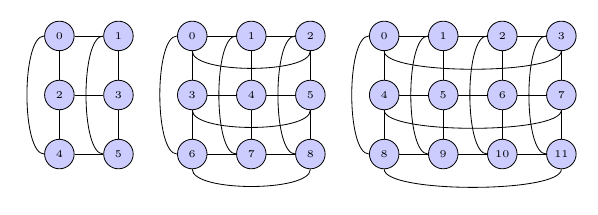}       
        };
    \end{tikzpicture}
    \begin{tikzpicture}
        \node at (-4, 2.5) {
            g) 
        };
        \node at (0,0) {
            \includegraphics{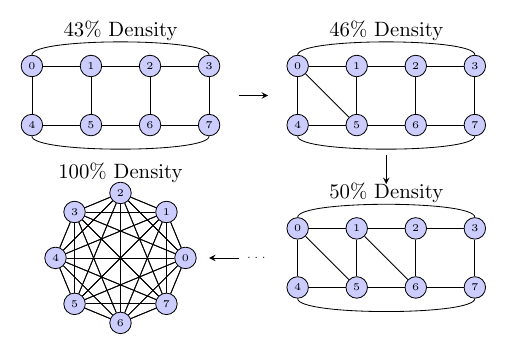}
        };
        \node at (5.5, 2.5) {
            h) 
        };
        \node at (8.8,0) {
            \includegraphics{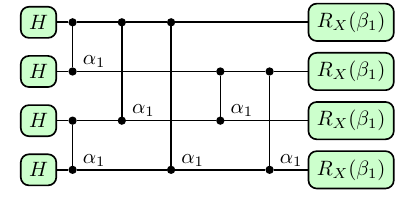}
        };
        \node at (12.5, 0.1) {
            $\rightarrow$
        };
        \node at (13.2,0.1) {
            \includegraphics{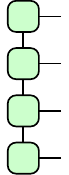}
        };
    \end{tikzpicture}
    \hspace{1cm}
    \caption{The various graph classes considered in this work are: a) Complete graphs with random edge weights (not shown), b) 3-regular graphs with random edge assignment and unit edge weights, c) 4-regular graphs with random edge assignment and unit edge weights, d)-f) grids of d) two, e) three and f) four columns with random edge weights, these are distinct from other 3 and 4-regular graphs as they have a specific edge structure. g) Beginning with a two-row grid with random edge weights, evolve by randomly assigning new edges with random weights selected from a uniform distribution over $[0,1]$ until a complete graph with $L\times 2$ nodes is formed. h) A single layer QAOA circuit for a four-node complete graph, with variational parameters $\alpha_1$ and $\beta_1$ as per Equation \eqref{eq:qaoa}. The QAOA circuit may be executed using an MPS simulator.}
    \label{fig:methods}
\end{figure*}
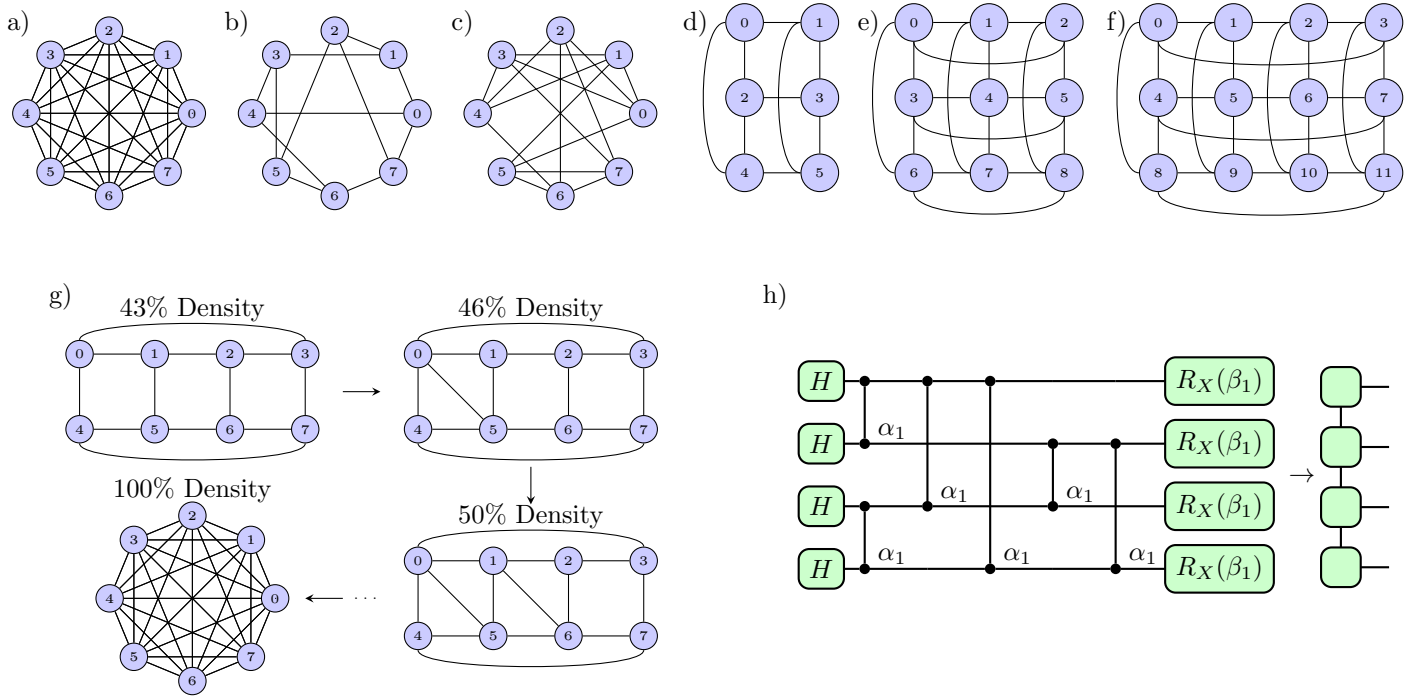

One of the most common combinatorial optimisation tasks studied using QAOA is the MAX-CUT problem\cite{farhi_quantum_2014,wang_noise-induced_2021,guerreschi2019qaoa,wang2018quantum,zhou2020quantum} which seeks to partition a graph $G$ with vertices $V$ and edges $E$ into two non-overlapping sets with the maximal number of edges between the two sets. The MAX-CUT problem can be represented by the following problem Hamiltonian\cite{lucas_ising_2014},
\begin{equation}
    H_P = \sum_{(i,j) \in E} w_{ij}Z_iZ_j, \label{eq:maxut}
\end{equation}
where $w_{ij}$ is the weight of the edge between nodes $i$ and $j$, and $Z$ is the Pauli $Z$ gate. This Hamiltonian may be solved using QAOA in a straightforward manner\cite{farhi_quantum_2014}. It has been established that a number of common optimisation problems may be formulated in terms of Hamiltonians of the form in Equation \eqref{eq:maxut} on specific graphs\cite{lucas_ising_2014}. For example, spin-glass models such as the Sherrington-Kirkpatrick model\cite{sherrington_solvable_1975} and Edwards–Anderson model\cite{SFEdwards_1975} can be reduced to MAX-CUT problems on complete and grid graphs respectively, both with edge weights $w_{ij}\in\{1,-1\}$. Additionally, MAX-CUT has applications in circuit design\cite{barahona1988application} where the underlying graph is planar, as well as machine scheduling\cite{alidaee19940} where the underlying graph is complete. 

In this work, complete graphs with random uniform weights $w_{ij} \in [0,1]$ and 3-regular graphs with unit weights, i.e. graphs with nodes each of degree three, will initially be considered similar to Ref. \cite{dupont_calibrating_2022}. Examples of such graphs can be seen in FIG. \ref{fig:methods}(a-b). We will be simulating these systems up to $p=6$ QAOA layers for all graphs, which is greater than the $p\leq2$ and $p\leq4$ simulations undertaken for complete and 3-regular graphs respectively in previous works\cite{dupont_calibrating_2022}. In doing so we seek to determine whether higher-depth QAOA has similar characteristics to low-depth QAOA. Such a distinction is important as in general one requires higher depth circuits in order to determine the solution state to high accuracy\cite{herrman_lower_2021}. The accuracy of QAOA may be measured using a common metric known as the approximation ratio which is introduced in Appendix \ref{apn_rat_def}. This metric is used to track the performance of the QAOA simulations for up to six layers for the types of graph considered in the main body of this work in Appendix \ref{apn_rat_res}.  

Additionally, grid graph simulations will be performed, as unlike random regular graphs this class has a structured edge layout and is planar making it similar to the underlying graphs of the common applications of MAX-CUT introduced above. Firstly, in order to draw comparisons between grid graphs and other graphs of the same uniform degree, 4-regular graphs with random uniform edge weights will also be simulated as per FIG. \ref{fig:methods}(c). Then, grids with two, three and four columns and random uniform edge weights will be considered as per FIG. \ref{fig:methods}(d-f). Simulation of all types of graphs considered in this work will be undertaken for $1,\dots,6$ layers. 

Another component of our study is to systematically explore the intermediate regime between grid graphs of two rows and complete graphs at two QAOA layers by starting with a grid graph and growing it by adding an edge with a randomly assigned edge weight between two nodes chosen at random that are otherwise not connected. This is repeated until the graph is complete. This process can be visualised in FIG. \ref{fig:methods}(g). We anticipate that the characteristics of the simulations, that is the entanglement entropy and simulation fidelity, will be dependent on the fundamental properties of the graph, that is the number of nodes, the number of edges and the layout of the edges. A metric that encapsulates the number of edges relative to the number of nodes is called the edge density and is defined on a graph $G = (V,E)$ as\cite{west2001introduction},
\begin{equation}
    \text{Density} = \frac{2|E|}{|V|(|V|-1)}.
\end{equation}
The density will be used as a point of comparison between different graphs when analysing how QAOA performs on these instances.

For all optimisations the SLSQP optimiser\cite{kraft_algorithm_1994} is utilised to minimise the QAOA parameters $\alpha_i$, $\beta_i$ with the following hyperparameters; \texttt{maxiter=500, ftol=1e-13, tol=1e-13}. The optimisation is repeated $10^3$ times with different initial $\alpha_i$, $\beta_i$ selected uniformly at random from $[-\pi, \pi]$. IBM's Qiskit \texttt{aer} simulator is used to find an initial set of optimal parameters, feeding this result into the \texttt{quimb} MPS simulator for the restricted entanglement simulations. Truncation of singular values to adhere to bond dimensions $\chi < 2^{N/2}$ is performed after each QAOA layer as per Ref. \cite{dupont_calibrating_2022}. For each type of graph defined, an average over 100 random instances is taken. That is in the case of fixed degree graphs, 100 different edge layouts, and in the case of complete graphs, 100 different realisations of the edge weights.

\subsection{Quantum Machine Learning}\label{sec:qml_method}
QNNs\cite{ballarin_entanglement_2023,kwak_quantum_2021}, sometimes referred to as QVCs\cite{west_benchmarking_2023,west_towards_2023} are machine learning models characterised by multiple layers of parameterised unitary transformations. They have found significant applications over recent years in image classification\cite{west_towards_2023,farhi_classification_2018}, drug response  predictions\cite{sagingalieva_hybrid_2023}, and breast cancer prediction\cite{li_model_2014}, amongst other tasks\cite{kwak_quantum_2021,glasser_probabilistic_2020,shi_approach_2020}. Unlike QAOA, QNNs have a flexible circuit structure, allowing for circuits to be structured in such a way that they may be efficiently executed on a quantum device\cite{leone_practical_2022}. In that sense they are similar to the Variational Quantum Eigensolver (VQE)\cite{kandala_hardware-efficient_2017}, with the primary distinguishing feature being that QNNs must first receive an input state which they will attempt to classify, as opposed to VQE and QAOA which simply optimise a given ansatz according to some cost function which completely defines the problem. One may also consider circuit ansätze which have a specified entanglement layout as defined by their two-qubit operators which is what will be considered in this work. In particular linear (both with and without cyclic boundary conditions) complete, single-qubit control and alternating-control layouts will be considered, each of which differs in their structure of CNOT gates. The circuits are shown in FIG. \ref{fig:qml-circs}. All ansätze have arbitrary parameterised single qubit rotations on all qubits. These single qubit rotations and entangling CNOT gates form a block that constitutes a single layer of the QNN which may then be repeated many times. The total number of gates for all the ansätze is primarily driven by the CNOT gates, with all but the full circuit ansatz having a gate count scaling of $O(pn)$ where $p$ is the number of layers and $n$ is the number of qubits. The full ansatz has a gate count scaling of $O(pn^2)$. The precise number of CNOT gates per layer is detailed in FIG. \ref{fig:qml-circs}.
\begin{figure*}
    \begin{tikzpicture}
        \node at (0,0) {
            \hspace{0.0em}\includegraphics[scale=0.385]{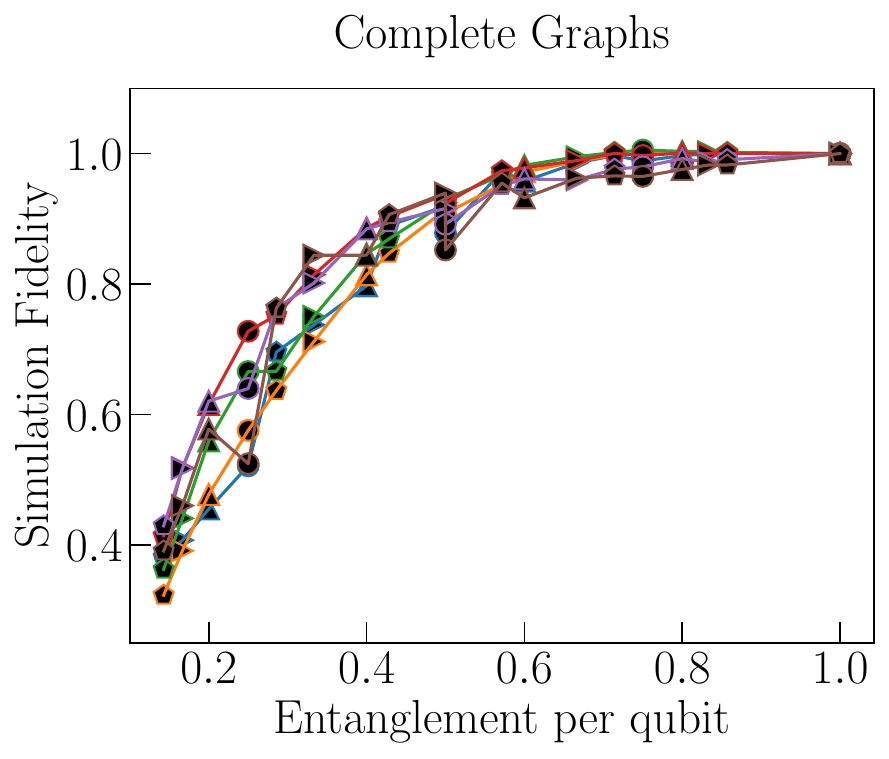}
        };
        \node at (5.75,0) {
            \includegraphics[scale=0.385]{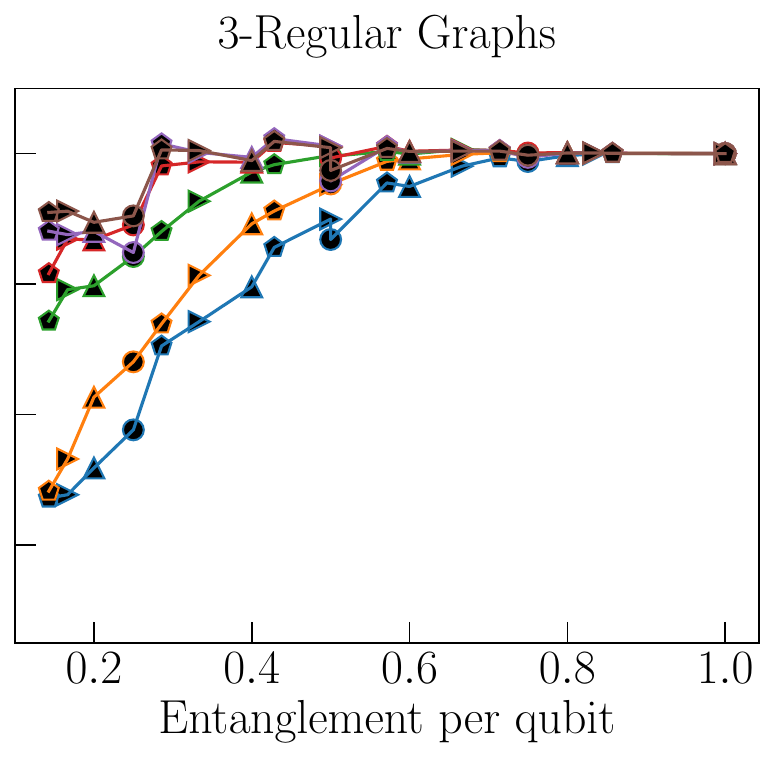}
        };
        \node at (12.2,0) {
            \includegraphics[scale=0.385]{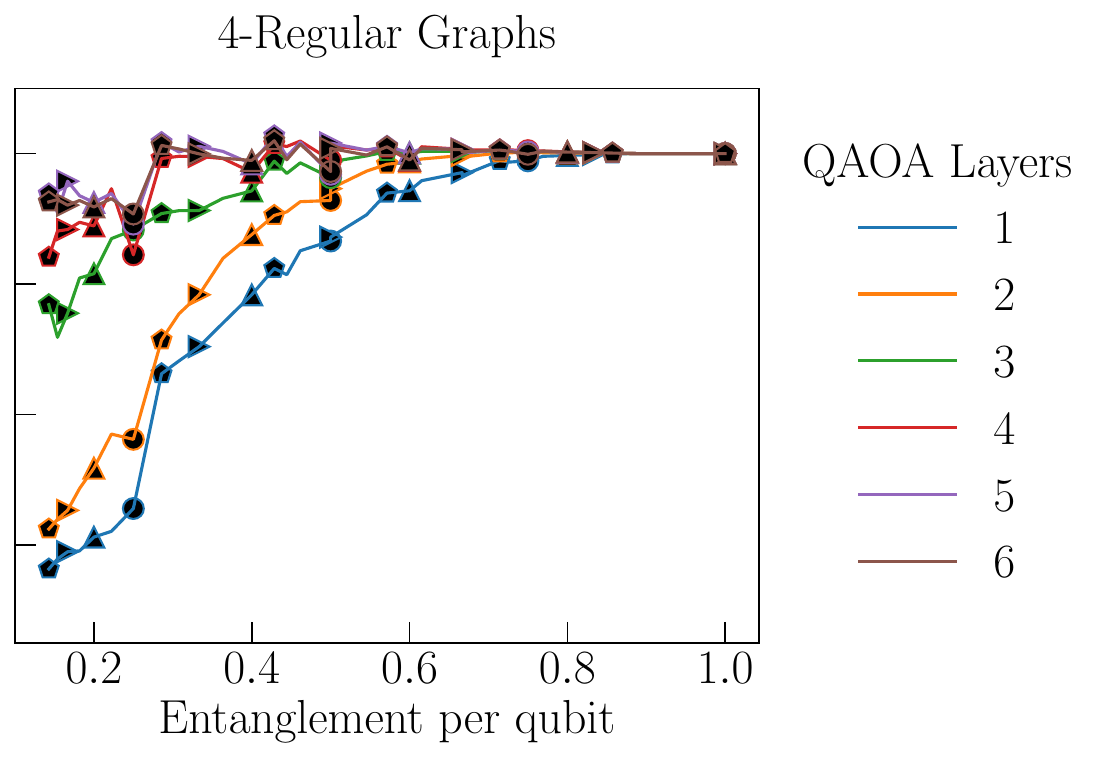}
        };
        \node at (0,-4.8) {
            \includegraphics[scale=0.385]{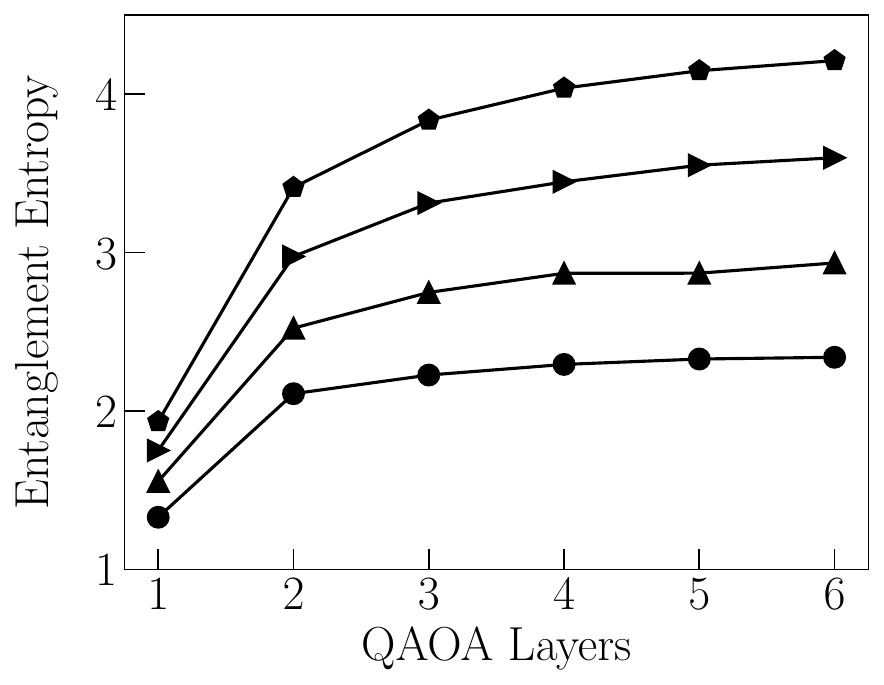}
        };
        \node at (5.75,-4.8) {
            \includegraphics[scale=0.385]{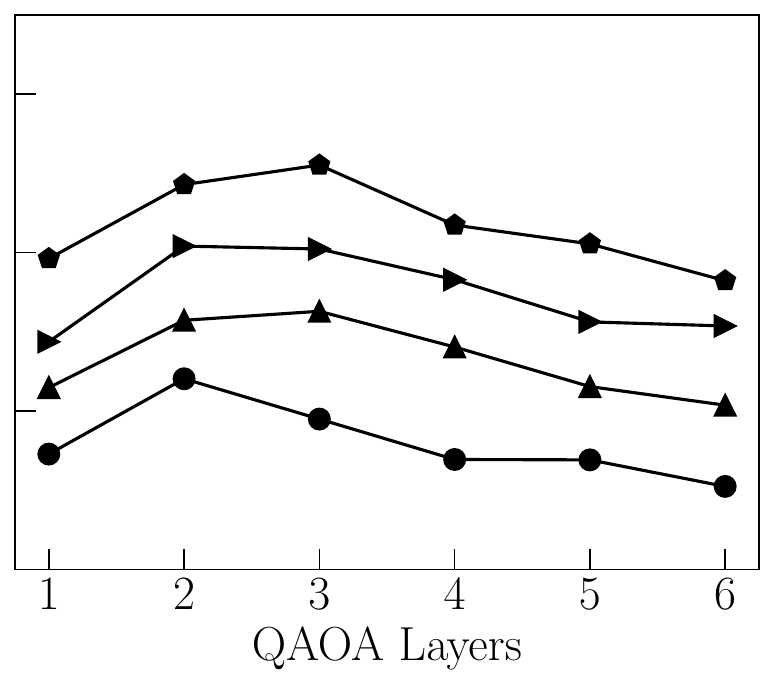}
        };
        \node at (12.1,-4.8) {
            \includegraphics[scale=0.385]{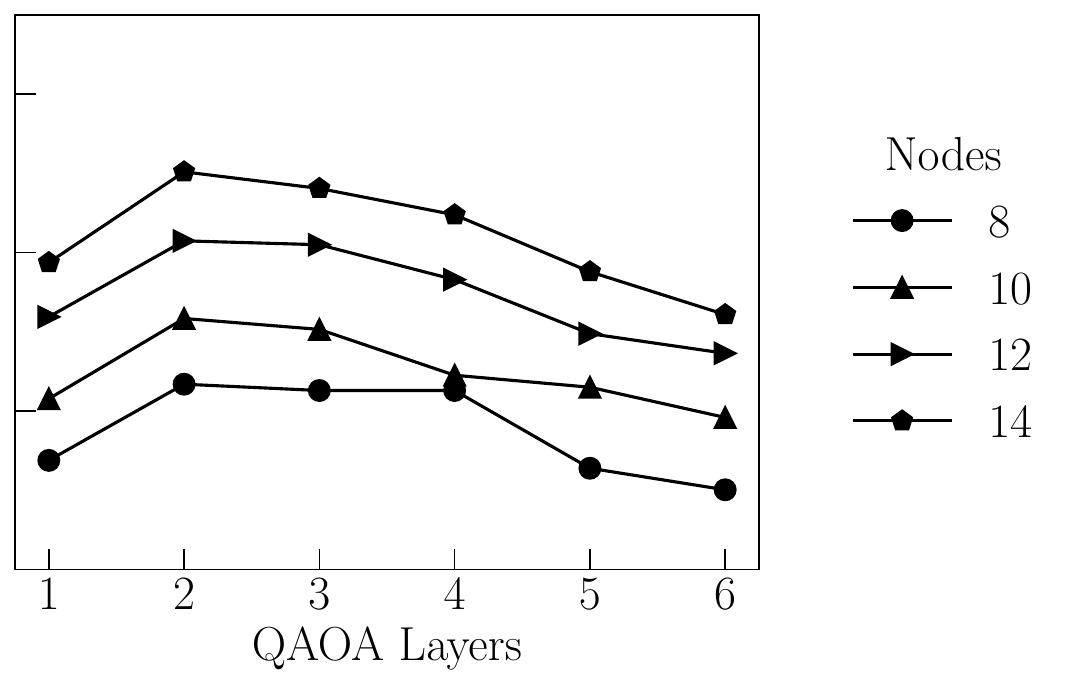}
        };
    \end{tikzpicture}
    \caption{The average simulation fidelity with restricted entanglement (top row), and average entanglement entropy at the end of the circuit (bottom row) of 100 random instances of complete, 3-regular and 4-regular graphs for up to six QAOA layers and up to 14 nodes. Note that after the peak entanglement entropy is reached for the regular graphs, the simulation fidelity with respect to the entanglement per qubit no longer follows a consistent scaling law as the number of graph nodes is varied. Additionally one can observe that the entanglement per qubit required to maintain high simulation fidelity decreases considerably.}
    \label{fig:reg}
\end{figure*}

We will be analysing QNNs which are trained to solve image classification problems, a task that has garnered particular attention in the QML community over recent years\cite{das_variational_2023,mathur_medical_2022,west_benchmarking_2023}. We will be using the MNIST\cite{li_deng_mnist_2012}, FMNIST\cite{xiao_fashion-mnist_2017} and CIFAR\cite{krizhevsky_learning_2009} datasets. The MNIST and FMNIST datasets are $28\times28$ greyscale images with 10 labels (or classes). The images are amplitude encoded\cite{larose_robust_2020} onto 10 qubits. Additionally, a simple binary classification task over the $0$ and $1$ classes of MNIST is performed. This is undertaken so that we have a model that can achieve very high test accuracy, ideally $100\%$. We call this modified dataset BMNIST. The CIFAR dataset contains $32\times 32$ RGB images with 10 classes which are amplitude encoded onto 12 qubits with the three colours encoded onto three separate channels. The task is reduced to a binary classification problem over the cars and boats classes in order to simplify training such that it can be performed on a classical simulator. For all datasets, we utilise 32,000 training examples and 1,000 test examples.

Classification is performed by assigning a qubit to each class and measuring the $Z$ expectation values of each qubit. The qubit with the largest $Z$ expectation value corresponds to the classification for the given input. In order to perform optimisation over this classification we assign a conditional probability that we find some class $c$ given some input $\mathbf{i}$ using softmax normalisation,
\begin{equation}
    p(c|\mathbf{i}) = \frac{\exp(\mel{\psi(\mathbf{i})}{Z_c}{\psi(\mathbf{i})})}{\sum_j\exp(\mel{\psi(\mathbf{i})}{Z_j}{\psi(\mathbf{i})})},
\end{equation}
which is fed into a cross entropy loss function which is minimised over. The \texttt{ADAM} optimiser\cite{kingma_adam_2014} with a learning rate of \texttt{5e-3} is utilised for training which is undertaken for a single epoch with a batch size of $16$. As per the QAOA simulations, truncation is performed after each layer, with normalisation of the state performed once at the end of the circuit.

\section{Results and Discussion}
In order to understand the role of entanglement in QAOA and QNNs, the entanglement entropy of each VQA is analysed throughout the execution of the algorithm with the untruncated ansatz. Additionally, an analysis of the fidelity of the simulation upon truncation is undertaken. These metrics provide an insight as to the role of entanglement in these algorithms and more specifically provide a benchmark that can be used to determine whether or not NISQ devices have sufficient capability to generate the entanglement required\cite{mooney_wholedevice_2021,hamilton_entanglement-based_2022}.
\begin{figure*}
    \begin{tikzpicture}
        \node at (0,0) {
            \includegraphics[scale=0.385]{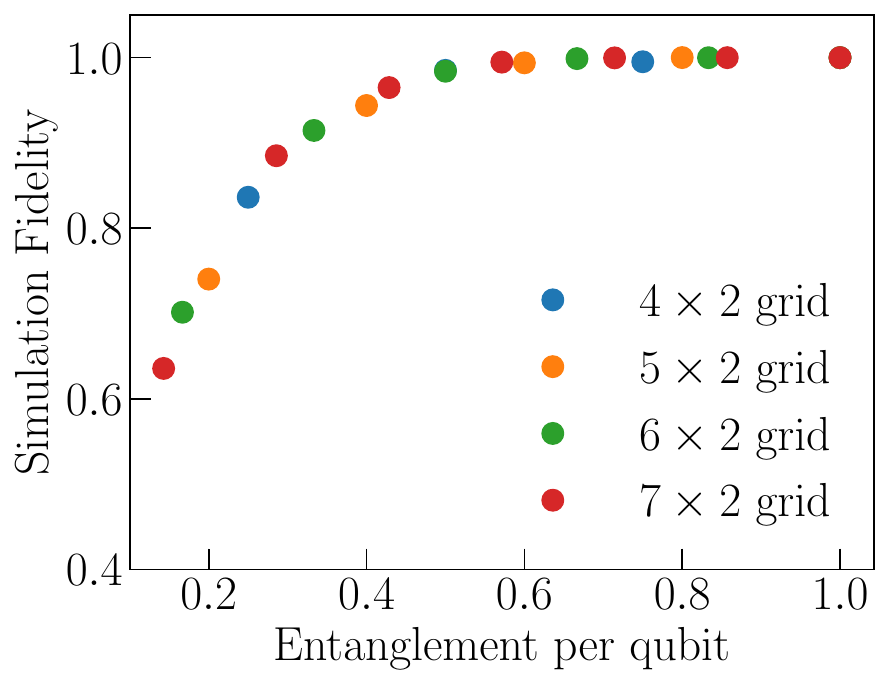}
        };
        \node at (5.75,0) {
            \includegraphics[scale=0.385]{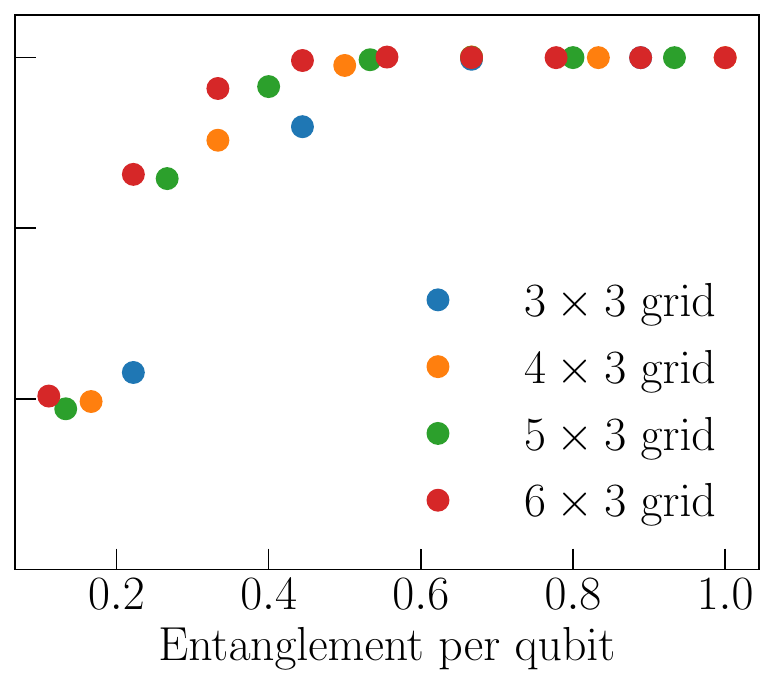}
        };
        \node at (11.2,0) {
            \includegraphics[scale=0.385]{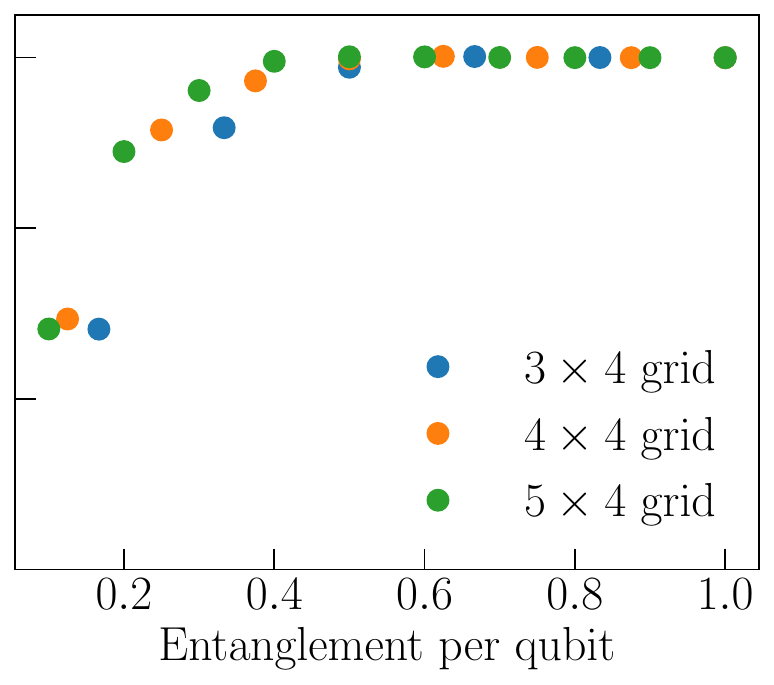}
        };
        \node at (0,-4.45) {
            \includegraphics[scale=0.385]{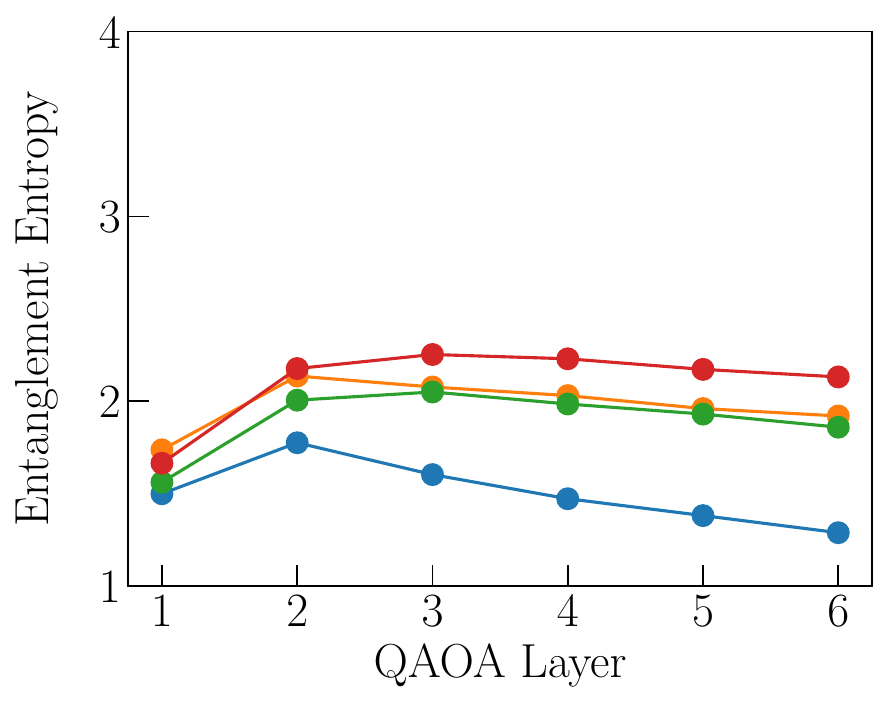}
        };
        \node at (5.75,-4.5) {
            \includegraphics[scale=0.385]{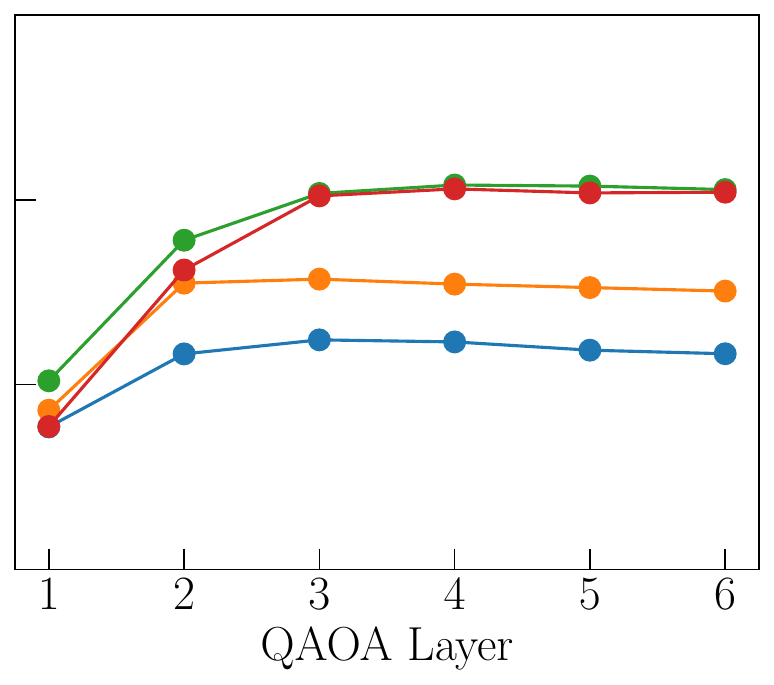}
        };
        \node at (11.2,-4.5) {
            \includegraphics[scale=0.385]{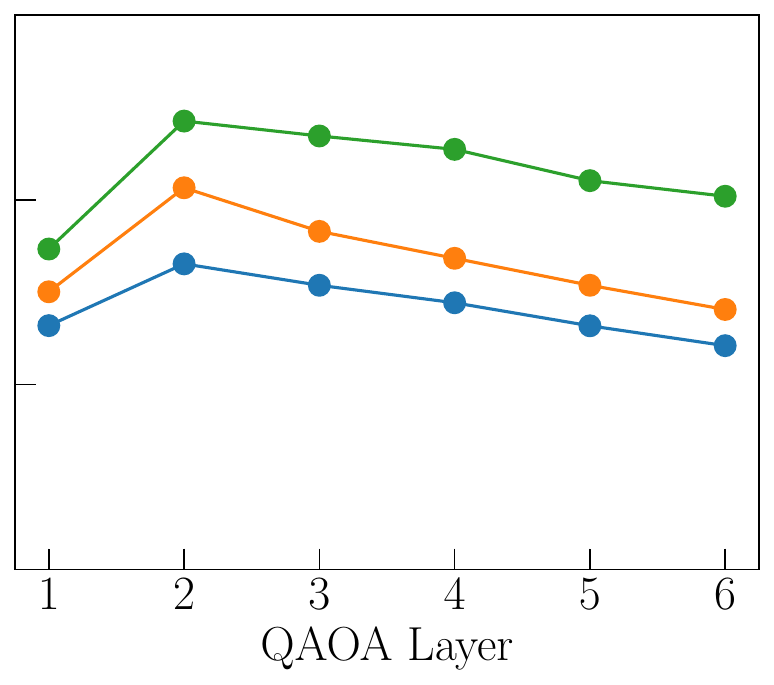}
        };
    \end{tikzpicture}
    \caption{The simulation fidelity at two QAOA layers (top row), and the entanglement entropy at the end of the circuit (bottom row) for grid graphs with two (left column), three (middle column) and four columns (right column) respectively. It is noted that despite the similarity to 4-regular graphs, grids of three and four columns appear to behave qualitatively differently from those of random 4-regular graphs. The simulation fidelity at four and six QAOA layers is presented in Appendix \ref{apn_big_qaoa}.}
    \label{fig:grids}
\end{figure*}

\subsection{Quantum Approximate Optimisation Algorithm}
QAOA on complete, 3-regular and 4-regular graphs as defined in Section \ref{sec:qaoa_method} is evaluated for up to six 
QAOA layers ($p\leq6$) with the number of nodes varying between $8$ to $14$. The fidelity of the restricted entanglement simulations and the entanglement entropy of the middle cut at the end of the circuit are shown in FIG. \ref{fig:reg}. It is found that for complete graphs and for 3-regular graphs at $p\leq2$, the simulation fidelity is consistent with the $\mathcal{F}(\ln(\chi)/N)$ scaling law established in Ref. \cite{dupont_calibrating_2022}. As the number of QAOA layers increases however we find that for 3-regular graphs the simulation fidelity remains consistently high as the entanglement per qubit is decreased. It is observed that in this case the simulation fidelity no longer follows a consistent scaling law as the number of nodes in the graph is varied. 

The entanglement entropy of the final state (that is at the end of the QAOA circuit) is also tracked for circuits of varying depth where it is found that entanglement entropy approaches a peak before starting to decrease for deeper circuits. Recognising that the solution to combinatorial optimisation tasks such as MAX-CUT are inherently product states, this can be interpreted as an indication that the QAOA circuit is preparing a state that is close to that of the desired ground state. Counter-intuitively this finding suggests that for such problems increasing the number of QAOA layers may result in a greater approximation ratio on devices incapable of creating high entanglement states.

Regarding 4-regular graphs, it is found that the entanglement entropy peaks at a lower depth, two QAOA layers for all $N\leq14$, compared to three layers for $N=10$ and $N=14$ in the 3-regular case. Furthermore, the scaling law that holds for two layers in the 3-regular case does not appear to be true for 4-regular graphs. This appears to be tied to the fact that the entanglement entropy reaches a peak with fewer QAOA layers. At higher depth, a similar trend to the 3-regular case is observed, where the state becomes more resilient to truncation of entanglement, with the overall entanglement entropy remaining low. Overall, the 4-regular case is found not to be significantly different to the 3-regular case. 

In the case of complete graphs, we find that the scaling law remains consistent as the number of QAOA layers is increased. Note that unlike the regular graphs considered above, there does not appear to be a peak in the entanglement entropy. This is consistent with our observations that the scaling law only holds below a certain number of QAOA layers where the entanglement entropy has not yet reached a maximum. Given the relatively lower approximation ratio for these states, as per Appendix \ref{apn_rat_res}, the monotonic increase in entanglement entropy is likely an indication that such graphs are more difficult for QAOA to solve rather than necessarily being an indicator that QAOA circuits for complete graphs exhibit inherently different characteristics. This is especially the case given that the true ground state is a product state, hence there must be a peak in entanglement entropy given that in the limit that the number of QAOA layers approaches infinity then the QAOA circuit can determine the ground state energy exactly\cite{farhi_quantum_2014}. Despite these observations, it is still not clear why the scaling law observed holds and how it is specifically related to the entanglement of the state. A more analytical study would need to be conducted to determine the theoretical basis for the behaviour of QAOA under these conditions.
\begin{figure}
    \includegraphics[scale=0.6]{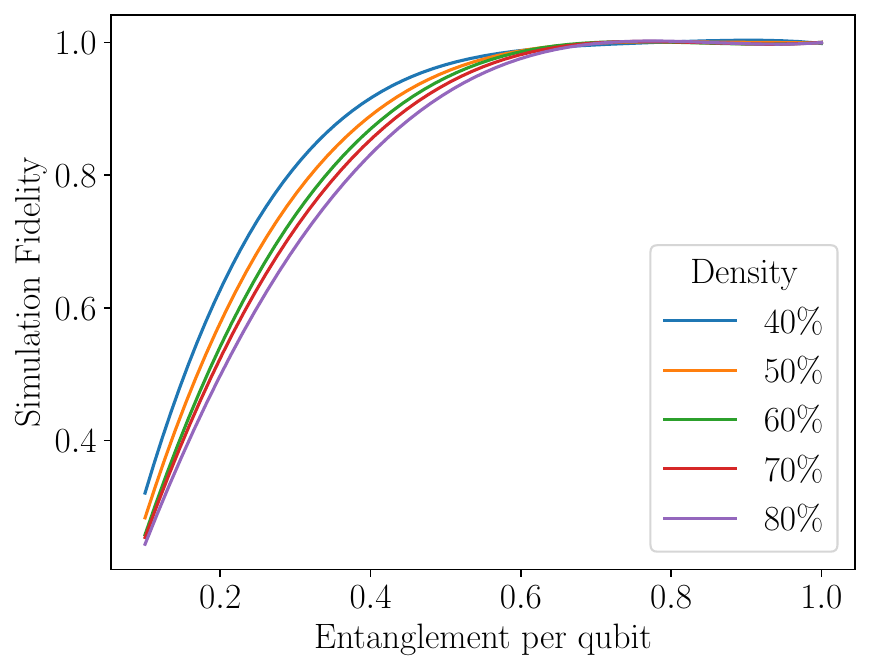}
    \caption{Best fit simulation fidelity for graphs with densities ranging from $49\%$ to $80\%$ at two QAOA layers. The number of nodes ranges from six to 16. It is apparent that for less dense graphs the resilience towards truncation is greater than that for graphs of higher density. For figures with the exact data points for each graph see Appendix \ref{apn_raw_den}.}
    \label{fig:trans}
\end{figure}

Investigating if imposing structure on the graphs has an effect on the simulation fidelity of QAOA, grid graphs are simulated using QAOA with results presented in FIG. \ref{fig:grids}. Results are split by number of columns in the grid graph. For the two-column case, our results closely resemble that of the 3-regular graph, noting that two-column (or equivalently two-row) grids are 3-regular. The primary distinction found is that the entanglement entropy remains low compared to similarly sized 3-regular graphs. We hypothesise that this is due to the fact that the edge layout in grid graphs is such that nodes that are situated far away from each other are in general less likely to be connected via an edge, given a node layout as per FIG. \ref{fig:grids}, which is the case for the simulations performed. This is compared to the random edges of random 3-regular graphs. Increasing the size to three and four-column grid graphs, which are now 4-regular, the results again appear similar to their random regular graph counterparts. The primary distinctions being as above, a decrease in entanglement entropy resulting from a node layout where nearby nodes are more likely to have edges compared to distant nodes.

These grid graph results highlight the importance of the mapping of nodes to qubits. Recognising in particular that by arranging nodes in such a way that their edges are preferentially between nodes that are near adjacent one can achieve a far lower entanglement entropy. This is of course not always possible. Continuing to consider grids whose nodes are labelled counting from left to right across each row, then top to bottom down each row, it is found that the exchange between rows and columns, in particular in the $3\times4$ case compared to the $4\times3$ case, appears to result in a significant difference in entanglement entropy along the middle cut. This is an indication that the entanglement is not evenly distributed for grid graphs, and hence node ordering may affect the optimisation step of QAOA, in particular in cases where entanglement is limited. Ultimately however for a given circuit with optimised parameters, this appears to only have a minimal effect on the simulation fidelity upon truncation. 

In addition to graph topology, graph density is another key characteristic of graphs that has an effect on the performance of QAOA circuits, with denser graphs resulting in circuits with a greater number of two-qubit gates and an overall greater depth. In order to investigate the effect of graph density, and specifically the transition from low-density graphs with structured edge layouts to high-density graphs, the procedure described in Section \ref{sec:qaoa_method} is followed to build a spectrum of graphs with varying density. The simulation fidelity for various graphs with densities ranging from $40\%$ to $80\%$ at two QAOA layers is reported in FIG. \ref{fig:trans}. It is found that the simulation fidelity decreases with the overall graph density, a result that appears consistent with that for the random regular and complete graphs. Likewise, there does not appear to be a universal scaling law for the simulation fidelity of the low-density graphs, with the scaling law only becoming more apparent as the density increases and one approaches a complete graph. This observation is more readily apparent in Appendix \ref{apn_raw_den} which presents the exact data points for graphs with $40\%$, $60\%$ and $80\%$ density.

Note in particular that the QAOA circuits are not trained using truncated MPS. This is because any further optimisation at a given maximal entanglement per qubit would simply exacerbate the observations of high simulation fidelity at lower entanglements per qubit which fundamentally does not alter the principal result of this work. Furthermore, it has been found that for one QAOA layer, the parameters corresponding to the minimum of an exact QAOA simulation are unchanged upon truncation \cite{dupont_calibrating_2022}. This consistency in the location of the minima has also been observed experimentally for single-layer QAOA evaluated on superconducting\cite{harrigan2021quantum,willsch2020benchmarking} trapped-ion\cite{pagano2020quantum} and photonic\cite{qiang2018large} quantum devices.

\subsection{Quantum Machine Learning}
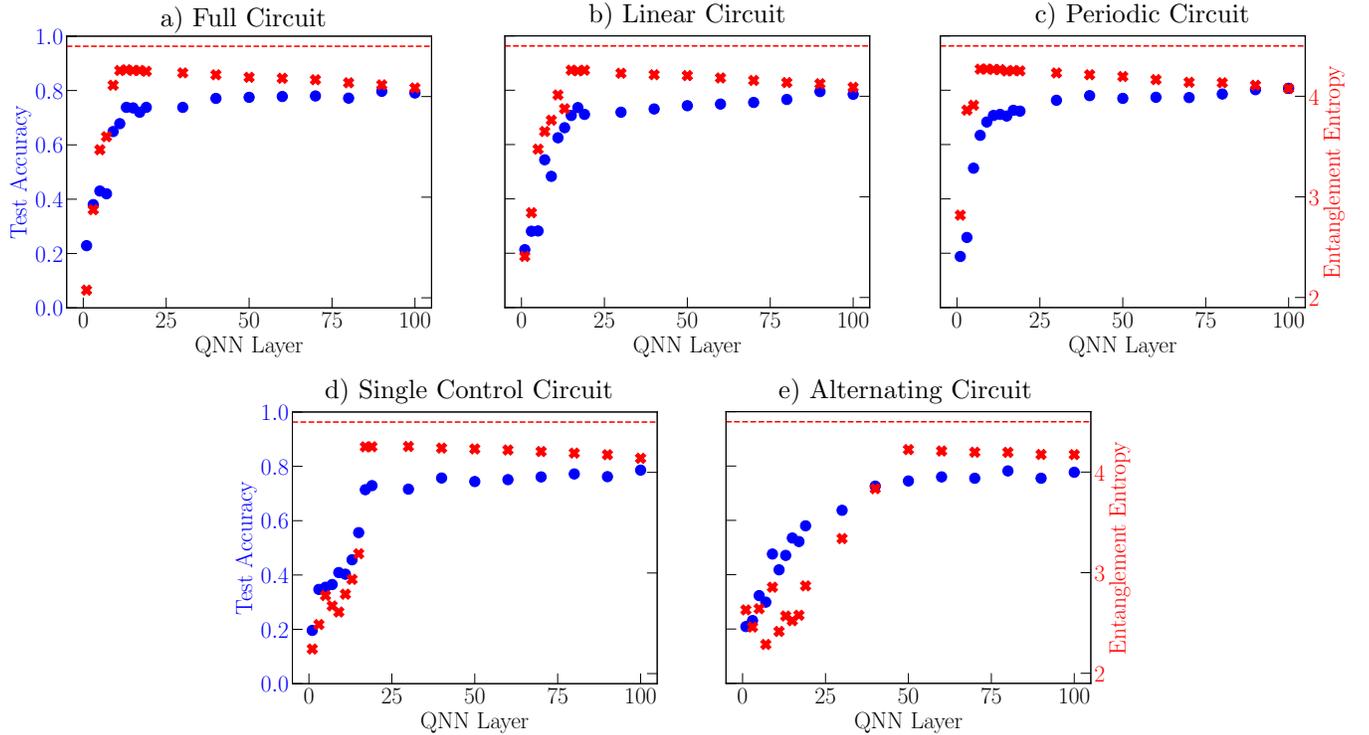
\begin{figure*}
    \begin{tikzpicture}
        \node at (0,0) {
            \includegraphics[scale=0.385]{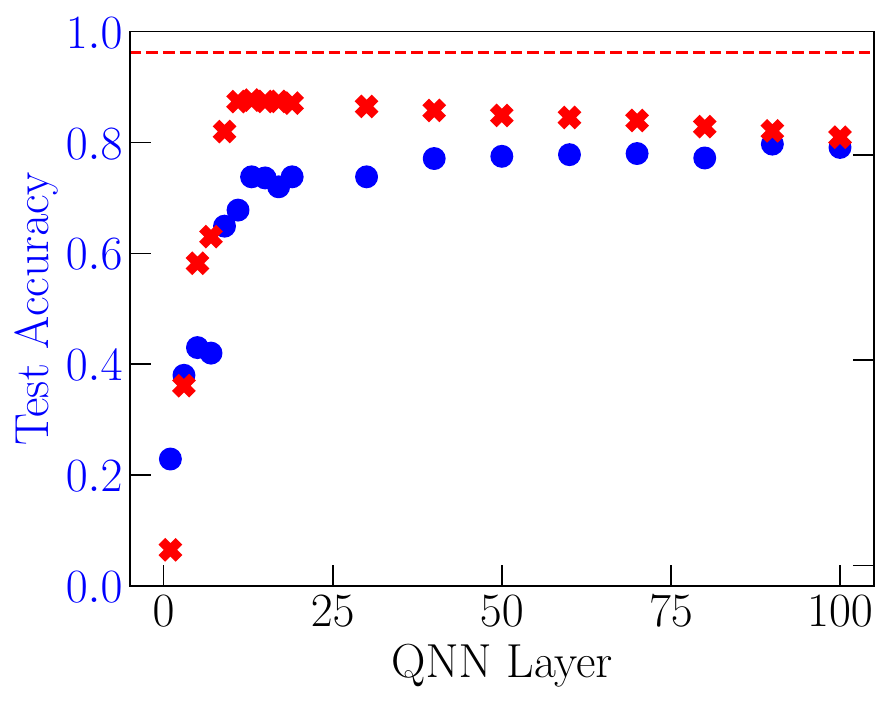}
        }; 
        \node at (0.3,2.3) {a) Full Circuit};
        \node at (6.2,-0.05) {
            \includegraphics[scale=0.385]{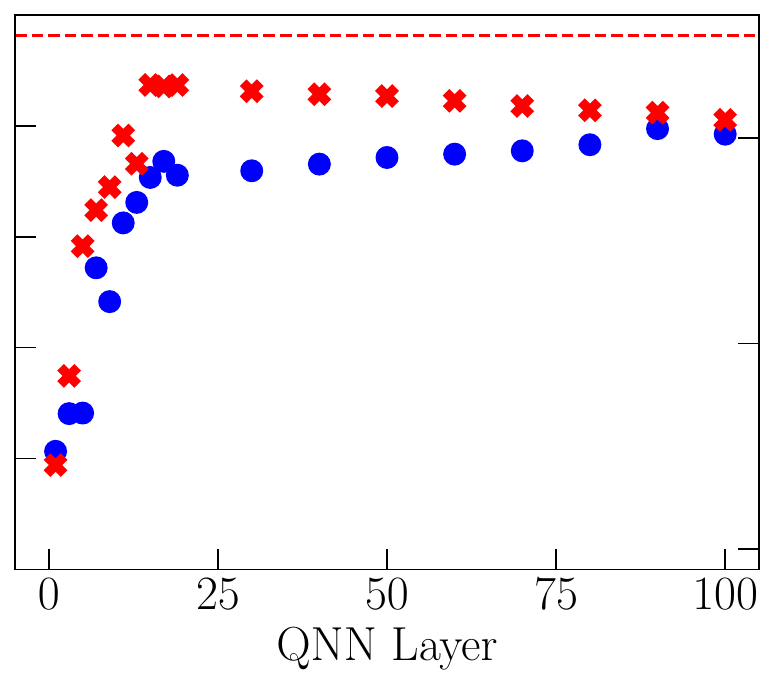}
        }; 
        \node at (6.2,2.35) {b) Linear Circuit};
        \node at (12.25,-0.05) {
            \includegraphics[scale=0.385]{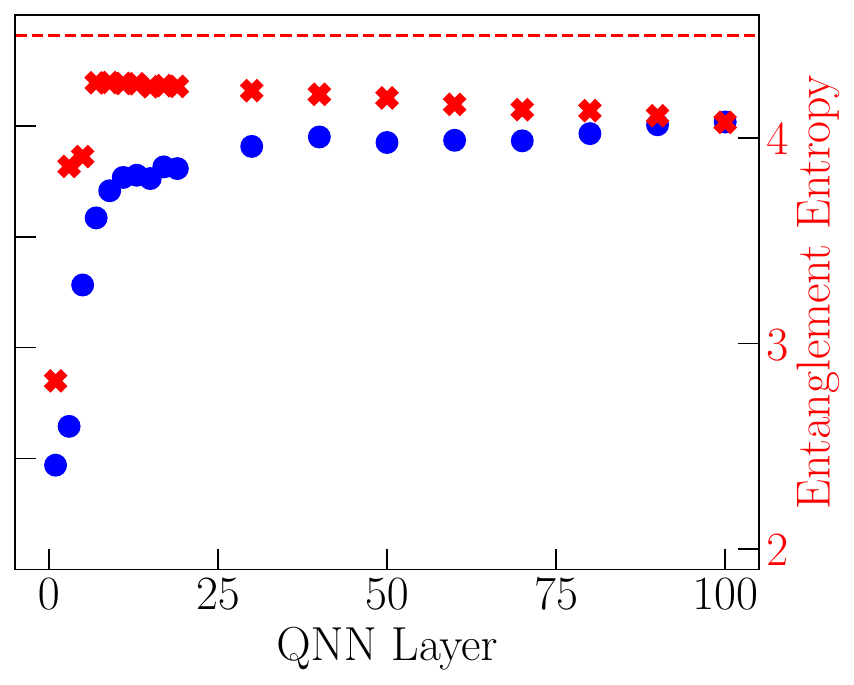}
        }; 
        \node at (12.25,2.35) {c) Periodic Circuit};
        \node at (3,-5) {
            \includegraphics[scale=0.385]{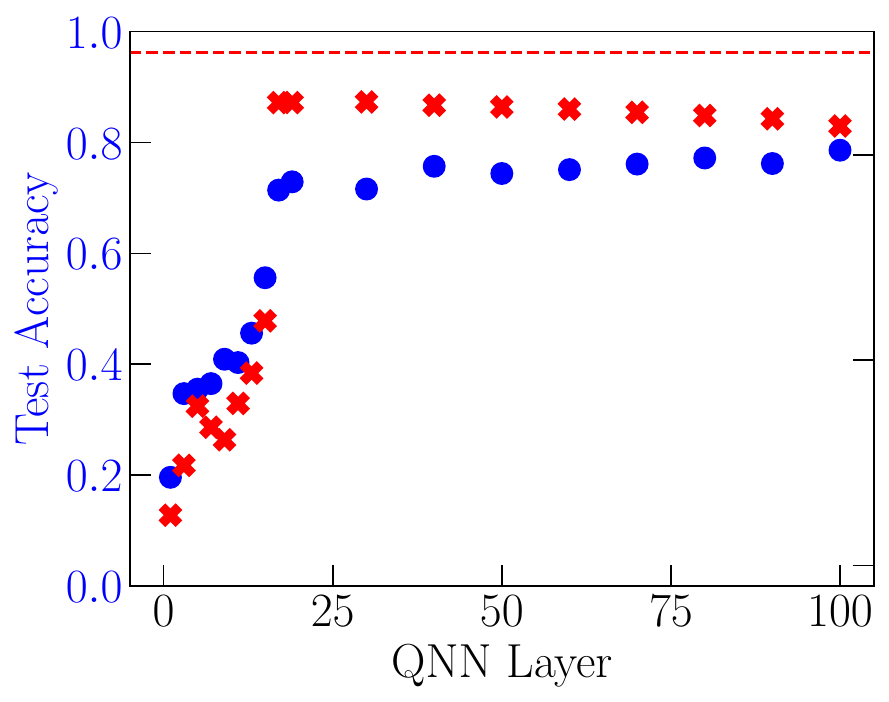}
        }; 
        \node at (3.3,-2.65) {d) Single Control Circuit};
        \node at (9.4,-5.05) {
            \includegraphics[scale=0.385]{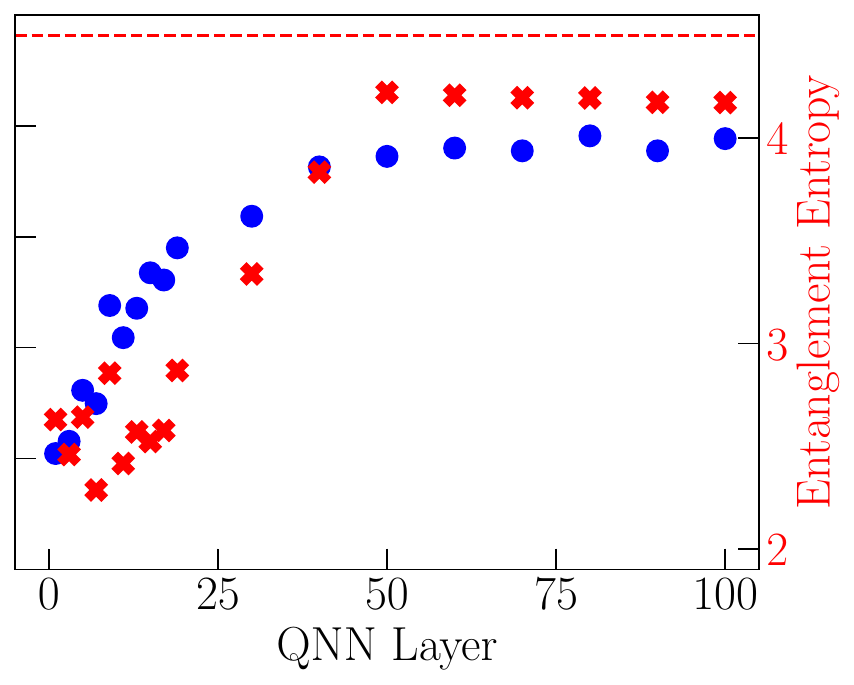}
        }; 
        \node at (9.1,-2.65) {e) Alternating Circuit};
    \end{tikzpicture}
    \caption{The test accuracy (blue dots) and average entanglement entropy (red crosses) at the end of the circuit for the circuits as per FIG. \ref{fig:qml-circs}. Each circuit is a $10$ qubit QNN with varying layers trained on the MNIST dataset. Note a rapid increase in test accuracy as the QNNs approach the maximum Haar entanglement entropy at $\approx 4.5$ (red dashed line)\cite{ballarin_entanglement_2023}. As the number of QNN layers increases beyond what is needed to reach the maximum entanglement entropy, a decrease in the entropy as the test accuracy steadily increases is observed. Note that the alternating ansatz in e) takes longer to reach the maximum entanglement entropy and performs worse than all other ansätze, which perform comparably.}
    \label{fig:circs}
\end{figure*}

The QNNs are trained on all datasets introduced in Section \ref{sec:qml_method} using the 5 different ansätze on a full state-vector simulator for up to 100 layers. The test accuracy and entanglement entropy of the final state for the various-sized QNNs trained on the MNIST dataset are shown in FIG. \ref{fig:circs}. Similar to QAOA it is found that the entanglement entropy at the end of the circuit increases as the number of QNN layers increases until a threshold entropy is reached, $<20$ layers for all but the alternating ansatz, after which the entanglement entropy appears to slightly decline with the test accuracy continuing to increase. We consider the structure of the alternating ansatz in Appendix \ref{apn_alt} and conclude that the Hilbert Space in which the state, prepared using one layer of the ansatz, lives is greatly reduced compared to the other ansätze considered in this work. As a result, it takes many more layers for the alternating ansatz to be able to produce near-maximally entangled states.

We note that a perfectly classified input would result in the final state of the circuit being a product state corresponding to the classification. This state would be of low entanglement. As a result, one can conclude that increasing the number of layers of the QNN results in a classification that is reproducible given the same input. This outcome is highly desirable. However, it is important to recognise that for systems where the number of qubits required for the input is equal to that required for the classification such a perfect outcome is not possible. This is because such an outcome would break the unitarity of the circuit, with in general many inputs leading to the same output. It is recognised that this is the case for the MNIST and FMNIST datasets which require $10$ qubits to be amplitude encoded and under a one-hot encoding scheme\cite{sammut2011encyclopedia} for classes will also require all of the same $10$ qubits for classification. Possible workarounds, if such a perfect classification is needed, are to separate the encoding and classification qubits or otherwise utilise another class encoding strategy that requires fewer qubits than is required for embedding the input.
\begin{figure}
    \begin{tikzpicture}
        \node at (0,0) {
            \includegraphics[scale=0.6]{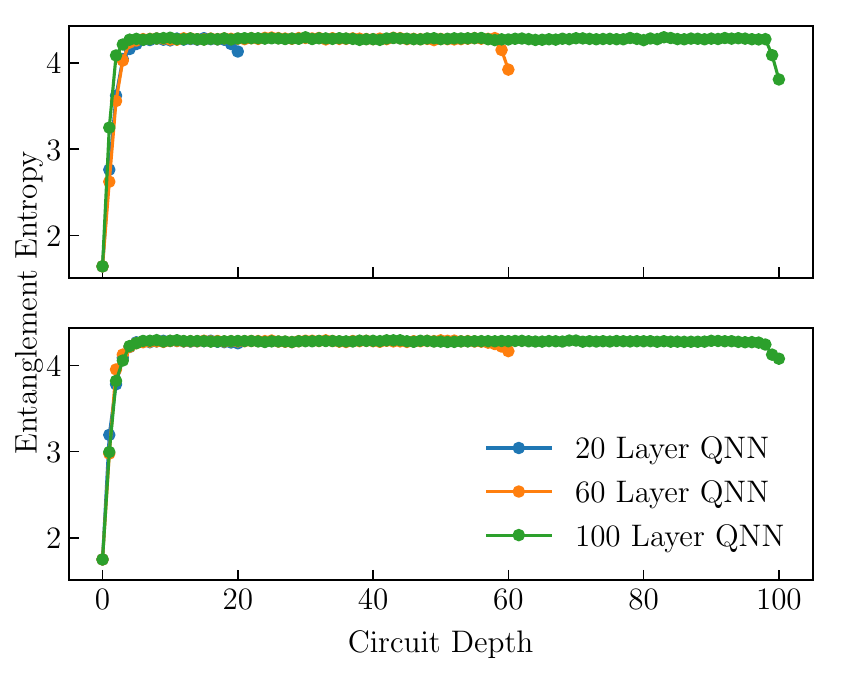}
        };
        \node at (-4.4,3) {a)};
        \node at (-4.4,-0.1) {b)};
    \end{tikzpicture}
    \caption{The evolution of entanglement throughout the circuit for 20, 60 and 100 layer periodic ansatz QNNs trained on a) the binary MNIST dataset and b) the $10$ class MNIST dataset. Test accuracies for a) are $0.998$, $0.998$ and $0.999$ respectively and $0.728$, $0.774$ and $0.807$ for b). Note that the entanglement entropy drops off towards the end of the circuit, with larger circuits showing a more significant drop in entanglement entropy towards the end of the circuit. The results for all other ansätze are provided in Appendix \ref{apn_qnn_entrop}.}
    \label{fig:2mnist-ent-growth}
\end{figure}

FIG. \ref{fig:circs} reveals a decrease in the entanglement entropy at the end of the circuit as the number of QNN layers is increased past a certain threshold. In order to hypothesise whether the QNNs will be resilient towards truncation, one must also consider the entanglement entropy throughout the evolution of the state. As such, the entanglement entropy throughout the execution of various-sized QNNs is tracked for the MNIST and binary MNIST datasets. The entropy at the end of each QNN layer for the periodic ansatz is shown in FIG. \ref{fig:2mnist-ent-growth}, with the remaining ansätze shown in Appendix \ref{apn_qnn_entrop}. It is observed that the decline in entanglement entropy occurs sharply at the end of the circuit and is not a gradual decrease similar to that seen in the analysis above. Hence, despite a decrease in the entanglement entropy of the final state, it is likely not the case that deeper circuits result in a greater resilience towards truncation similar to that which was observed for QAOA on regular graphs. This observation would also suggest that training a QNN layer by layer is unlikely to be successful as the final state of some QNN with $a$ layers appears to be substantially different to a QNN with $b>a$ layers at layer $a$. This is in contrast to QAOA, where it is possible to see improvement in the cost by training a system layer by layer up until $N$ total layers where $N$ is the number of qubits\cite{PhysRevA.104.L030401}.

\begin{figure}
    \begin{tikzpicture}
        \node at (2.4,2.2) {
            \includegraphics[scale=0.3]{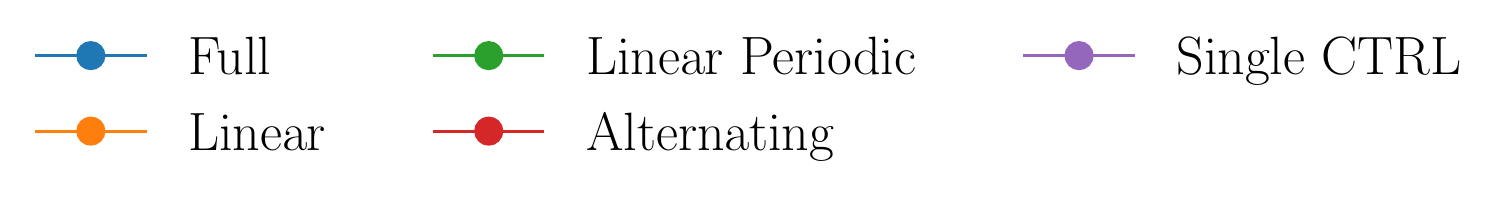}
        };
        \node at (0,0) {
            \includegraphics[scale=0.3]{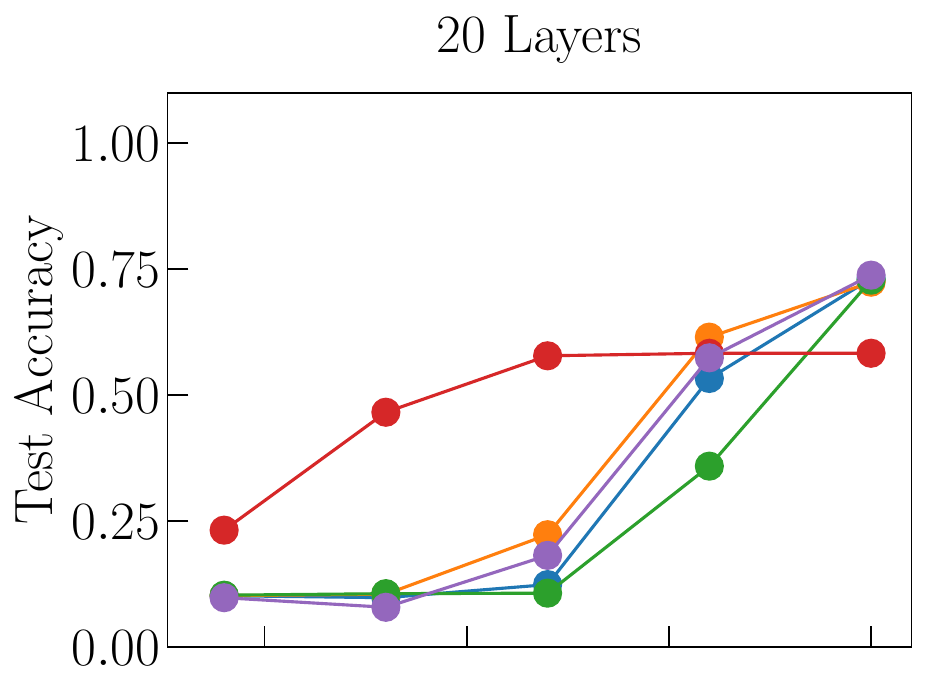}
        };
        \node at (4.75,0.053) {
            \includegraphics[scale=0.3]{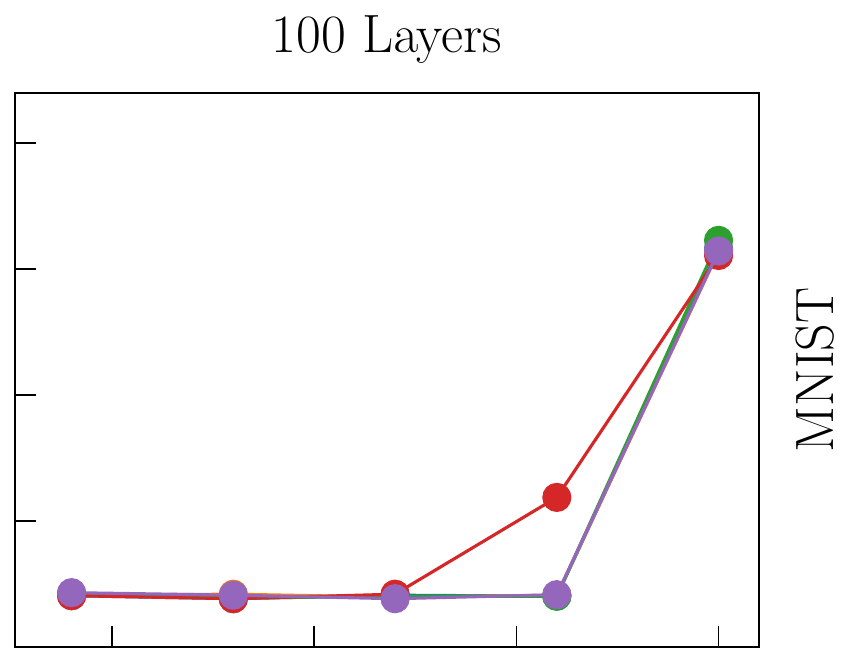}
        };
        \node at (0,-3.2) {
            \includegraphics[scale=0.3]{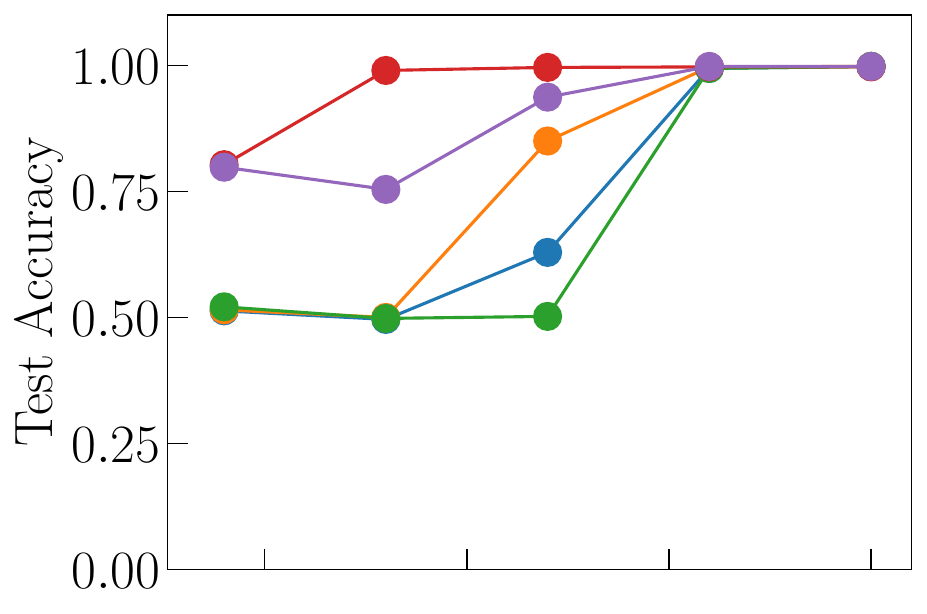}
        };
        \node at (4.75,-3.15) {
            \includegraphics[scale=0.3]{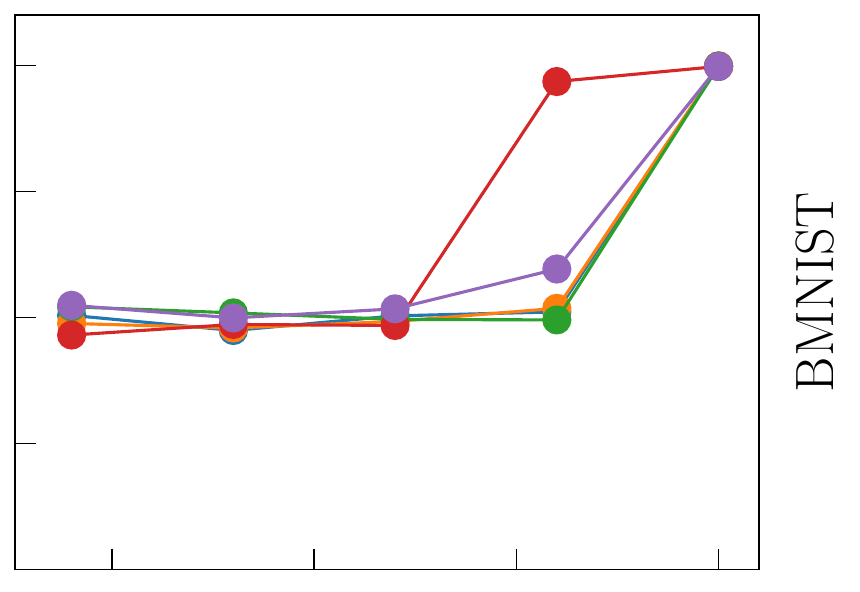}
        };
        \node at (0,-6.425) {
            \includegraphics[scale=0.3]{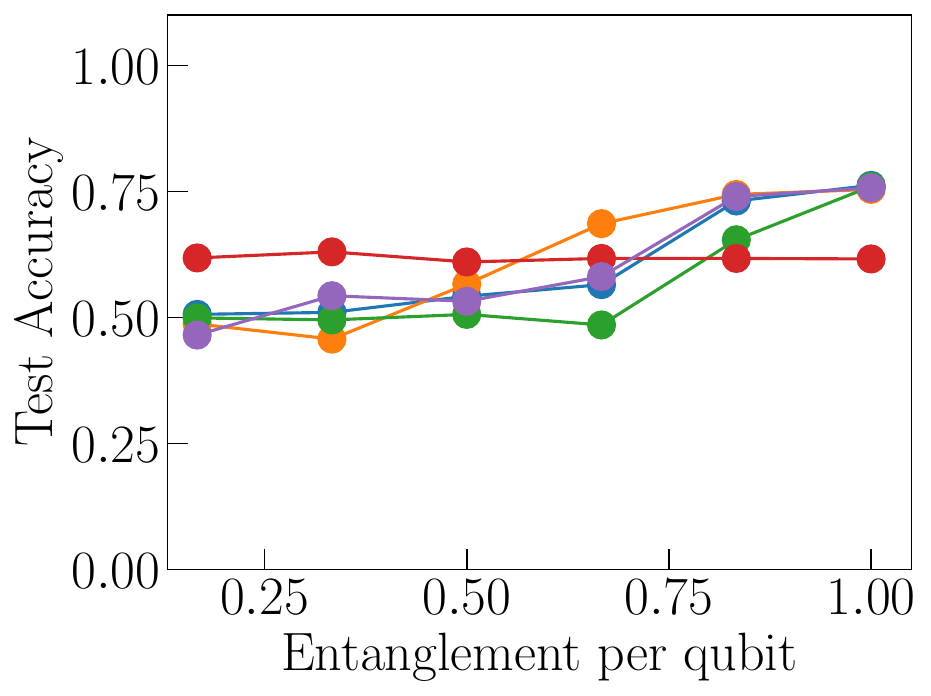}
        };
        \node at (4.75,-6.425) {
            \includegraphics[scale=0.3]{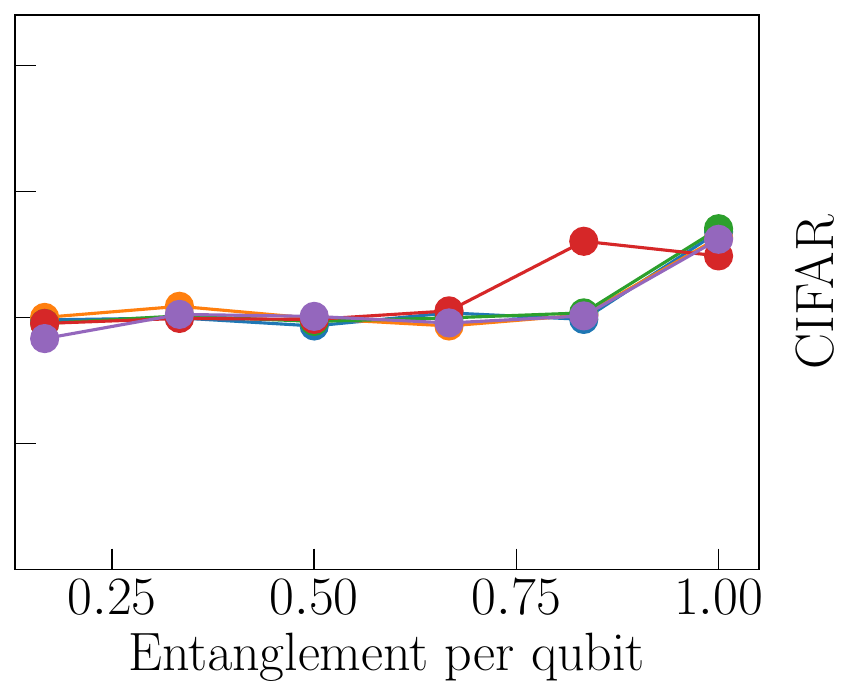}
        };
    \end{tikzpicture}
    \caption{Test accuracy for various datasets under limited entanglement simulations. Note that deeper QNNs are more able to utilise entanglement in a useful manner. We recognise that this is not necessarily a result of shallower circuits being less entangled as shallower circuits are found to have as high an entanglement entropy for QNNs above $20$ layers. In general, for the MNSIT binary classification task, there appears to be no reason that one needs to extend beyond a shallow, low entanglement circuit. For the binary CIFAR classification task it is observed that there is a notable decrease in the test accuracy for deeper QNNs, this is possibly a result of entanglement-induced barren plateaus. The results for 60 layers and the FMNIST dataset can be found in Appendix \ref{apn_qnn}.}
    \label{fig:test_accs}
\end{figure}

To test this hypothesis, MPS simulations with truncation are run across the entire test set with test accuracy for each entanglement per qubit shown in FIG. \ref{fig:test_accs} and Appendix \ref{apn_qnn}. For the monochromatic $28\times28$ pixel MNIST and FMNIST datasets, it is found that all but the alternating ansatz appear to not be resilient to any truncation at $20$, $60$ or $100$ QNN layers. This is in agreement with the observation made above that resilience towards truncation is unlikely even when the final state is of relatively lower entanglement. Counter-intuitively this means that provided a trained QNN and a quantum device or simulator incapable of achieving high entanglement, it may be more beneficial to use the alternating ansatz at low depth as it is capable of achieving higher test accuracies with less entanglement compared to the other ansätze. As the number of QNN layers is increased however this advantage disappears with all ansätze not showing any resilience towards truncation. Additionally, factoring in the gate requirements of each ansatz as per TABLE \ref{tab:gates}, it is noted that despite the vastly greater number of gates the full circuit ansatz does not appear to perform substantially better than the ansätze which only have a CNOT gate count which scales linearly with the number of qubits per layer. 
\begin{table}[]
    \centering
    \begin{tabular}{|c|c|c|c|c|}
        \hline
        $p$ & Circuit  & Accuracy & \#Gates & $S$ \\
        \hline
         20 & Full & 0.731 & 1,100 & 2.254\\
          & Linear & 0.724 & 400 & 4.242 \\
          & Periodic & 0.728 & 420 & 4.256 \\
          & Single Control & 0.738 & 400 & 4.251 \\
          & Alternating & 0.583 & 400 & 2.673 \\
         \hline 
         60 & Full & 0.778 & 3,300 &  4.182 \\
          & Linear & 0.749 & 1,200 & 4.182 \\
          & Periodic & 0.774 & 1,260 & 4.165 \\
          & Single Control & 0.751 & 1,200 & 4.222 \\
          & Alternating & 0.760 & 1,200 & 4.209\\
         \hline 
         100 & Full & 0.791 & 5,500 & 4.086 \\
          & Linear & 0.785 & 2,000 &4.089\\
          & Periodic & 0.807 & 2,100 & 4.077 \\
          & Single Control & 0.786 & 2,000 & 4.140 \\
          & Alternating & 0.777 & 2,000 & 4.173 \\
        \hline
    \end{tabular}
    \caption{The test accuracy, number of gates and entanglement entropy, $S$, of the different QNN circuit ansätze trained on the MNIST dataset at $p=20$, $60$ and $100$ layers. Despite requiring substantially more gates, the fully connected ansatz does not result in any substantial improvement in the test accuracy nor is the entanglement entropy substantially different to the other ansätze, except the alternating ansatz.}
    \label{tab:gates}
\end{table}

For the simple binary MNIST dataset, it is found that the alternating and single control ansätze show a significant resilience towards truncation at $20$ QNN layers, and to a lesser extent at $60$ QNN layers. All other ansätze show some resilience at $20$ layers and $60$ layers, but in general, appear to have a lower test accuracy for a given entanglement per qubit compared to the single control and alternating circuits. It is noted however that the entanglement entropy in FIG. \ref{fig:2mnist-ent-growth}(a) continues to decrease for QNNs with a greater number of layers. This is not inconsistent with the findings here of high test accuracy at low depth, noting in particular that classification is performed by taking the largest $Z$ expectation value taken over many shots. Hence very high test accuracy does not necessarily correspond to a very high probability of measuring the output corresponding to the correct classification. Despite this, as the primary goal of QNNs is accurate classification it remains beneficial to use circuits of low depth that are resilient towards truncation which are able to achieve high classification accuracy. For simple datasets such as binary MNIST, this can be achieved with $<20$ layers with the alternating ansatz.

Analysing the more complex $32\times32$ RGB CIFAR dataset, it is found that the resilience towards truncation as circuit depth increases follows a similar pattern to the greyscale datasets. Interestingly, however, it is found that for the $20$ layer QNN, the alternating ansatz maintains a complete resilience towards truncation, albeit at the cost of a reduced test accuracy compared to the other ansätze. We hypothesise that this is because the alternating ansatz is encoding an exceedingly simple model. However, as the depth increases a more sophisticated classification is performed and this resilience is lost. At this high depth there also appears to be a slight decrease in the overall test accuracy for all but the alternating circuit. It is possible this is a result of entanglement-induced barren plateaus\cite{mcclean2018barren,PRXQuantum.2.040316}, however, one would need to determine whether the entanglement entropy of the ansätze satisfies a volume law for this to be the case.

Additionally, at high depth, there is a minor increase in the test accuracy as we perform moderate truncation for the alternating ansatz. Further investigation is required to determine the underlying mechanism here, but we hypothesise that the act of truncation, which is inherently a non-unitary evolution of the state, is such that the system has access to a part of the Hilbert space that was not available to it with just the standard unitary operations of the circuit ansatz. A similar phenomenon is observed for QAOA on grid graphs at higher depths as discussed in in Appendix \ref{apn_big_qaoa}. Additionally, an overall decrease in the test accuracy across all ansätze for high-depth circuits is observed, which is likely due to entanglement-induced barren plateaus. 
 
Like QAOA, the QNNs are not trained using truncated MPS simulations, meaning that the simulations performed are assessing the scenario where one is capable of training the model with a highly entangled device but must use a low entanglement device to evaluate the model on new inputs. We hypothesise that training the QNNs with a truncated MPS simulation may result in a marked improvement in the overall test accuracy. This follows from Ref. \cite{west_approximate_nodate} where the authors found that training QNNs with noisy inputs results in a notable improvement in test accuracy when tested on similarly noisy inputs. This is compared to QNNs that are trained using exact inputs but tested with noisy inputs where they found that the test accuracy was reduced. As the truncated MPS simulations are significantly slower than training using \texttt{pennylane} this hypothesis was not tested as it was not feasible to perform adequate training with the computational resources available to us.

\subsection{Summary of comparison between QAOA and QNN}

Our work has revealed that the entanglement of VQAs is highly dependent on the structure of the problem. In general, we find that for combinatorial optimisation problems solved using QAOA, high-depth circuits do not necessarily result in higher entanglement in circuits, with many problem instances showing some resilience towards truncation. This is compared to image classification tasks solved using QNNs, where we find that those circuits are much more entangled at high depth, and are far less resilient to any truncation of entanglement. Our results suggest that  compared to QNN, generally QAOA is more likely to be within the capability of NISQ devices due to lower demand on the entanglement in the circuits. However, for some simple datasets such as binary MNIST, QNN models using alternating ansatz may also be within the reach of near-term devices due to relatively low-depth circuits requiring low entanglement. For sufficiently complex datasets, however, it is found that higher-depth and highly entangled QNNs are required to maintain high test accuracies. 

\section{Conclusion}
Variational quantum algorithms show significant promise in the NISQ era of quantum computing as a result of their ability to perform classically difficult tasks using relatively few qubits and circuits of shallow depth. In this work, we find that two of the leading VQCs, QAOA and QNNs, show a varying degree of resilience towards the truncation of entanglement. Importantly though, we find that the resilience of any given circuit is highly dependent on the depth of the circuit and the variational ansatz, recognising that different problems have inherently different requirements regarding circuit depth and structure. As a result, a previously observed scaling law\cite{dupont_calibrating_2022} does not necessarily hold when considering specific problem instances. 

For QAOA, we relate the simulation fidelity to the entanglement entropy of the final states produced, finding that monotonic scaling laws with respect to the size of the system for each type of graph being solved are more likely to hold in the regime where the entanglement entropy has not yet reached a peak. We also find that structured regular graphs, such as the grid graph appear to exhibit scaling relations that are inherently distinct from random regular graphs, ultimately finding that the scaling relation approaches that of the complete graph upon increasing the density when interpolating from an initial grid graph. This has implications for problems such as the vehicle routing problem map onto low-density graphs of a given structure. 

It still remains an open question as to the underlying mechanics behind the scaling relations observed. However one can relate the eventual increase in resilience to truncation to the simple nature of the final solution state of combinatorial optimisation problems being a product state. We recognise that any of the combinatorial optimisation problems considered can be solved exactly with a trivial product ansatz with $3N$ parameters, where $N$ is the number of qubits, but it remains open as to how efficiently such an optimisation is compared to a QAOA ansatz with fewer variational parameters. 

For QNNs, we find that all but the simplest datasets require highly entangled ansätze in order to achieve high test accuracy, with high-depth circuits exhibiting higher test accuracies and low resilience towards truncation of entanglement. We find that, unlike QAOA, despite a decrease in entanglement entropy for higher-depth models, resilience towards truncation decreases. For more complicated datasets such as the RGB CIFAR dataset, we sometimes observe a slight increase in test accuracy upon truncation, which is similar to an observation made for QAOA on certain problems at high depths. It is not clear what the underlying cause of this is, and as such this remains an open question. Given the relatively low test accuracies at which this phenomenon occurred, however, it is not clear whether or not this would continue to hold for models of interest that achieve a high test accuracy.

Overall, our results based on a systematic set of simulations have provided important new insights into the role of entanglement in the performance of two widely used variational quantum algorithms. In the current NISQ era of quantum computing where noise or errors in quantum devices limit the entanglement that can be generated in quantum circuits, our work will enable their implementations with optimal accuracy within the constraint of affordable entanglement. Future avenues of research include the investigation of other VQAs such as those used to simulate spin systems or those used in quantum chemistry.

\section*{Acknowledgements}
The authors acknowledge useful discussions with Charles D. Hill in the early stages of the project, and with Maxwell T West on the QNN part of the work. ACN acknowledges the support of the Australian Government Research Training Program Scholarship. The authors acknowledge the support from IBM Quantum Hub at the University of Melbourne. The computational resources were provided by the National Computing Infrastructure (NCI) and Pawsey Supercomputing Center through the National Computational Merit Allocation Scheme (NCMAS).
\\ \\
\section*{Code Availability}
The code used to generate all results in this paper is available upon reasonable request.

\bibliographystyle{naturemag}
\bibliography{references}

\cleardoublepage
\appendix
\section{A note on the Simulation Fidelity for QAOA} \label{apn_sim_fid}

The simulation fidelity for QAOA may be explicitly written down as,
\begin{equation}
    \text{Simulation Fidelity} = \frac{\mel{\psi_\chi}{H_P}{\psi_\chi}}{\mel{\psi}{H_P}{\psi}}, \label{eq:apen_sim}
\end{equation}
where $\ket{\psi_\chi}$ is an approximation of state $\ket{\psi}$ with bond dimension $\chi$. This definition, similar to that of the approximation ratio suffers from a certain ambiguity. This is discussed in greater depth in Appendix \ref{apn_rat_def}. We note however that for all instances considered in this work we have $\mel{\psi}{H_P}{\psi} < 0$. To illustrate this we provide the average $\mel{\psi}{H_P}{\psi}$ for the complete, 3-regular, 4-regular instance in this work in FIG. \ref{fig:costs}. Despite the issues with the definition in Equation \eqref{eq:apen_sim}, we continue to use it so that our findings remain directly comparable to previous works\cite{dupont_calibrating_2022} which we compare our results against.
\begin{figure}
    \begin{tikzpicture}
        \node at (0,0) {
            \includegraphics[scale=0.32]{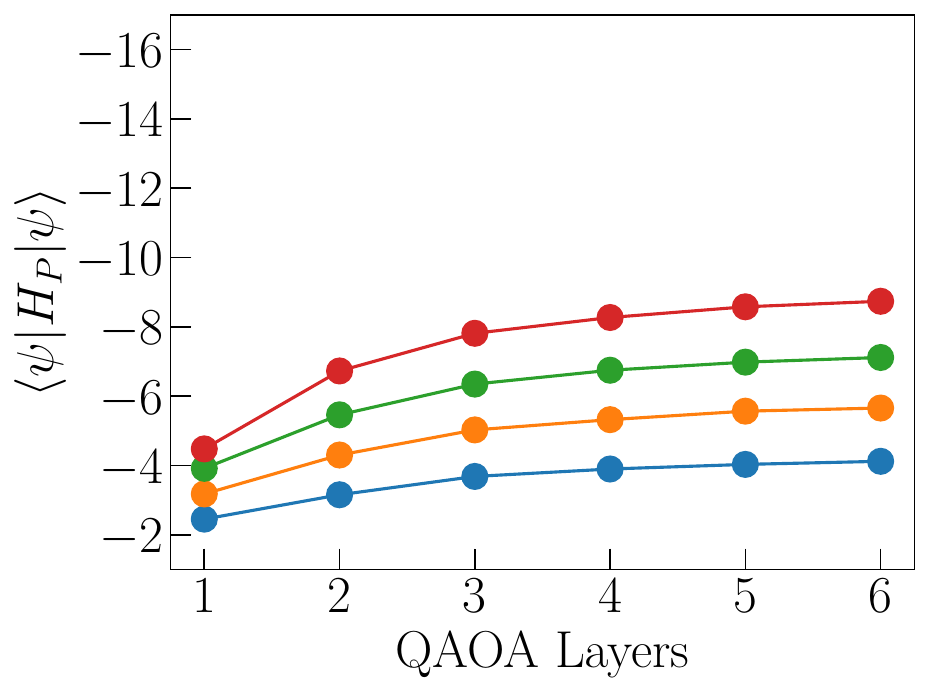}
        };
        \node at (-1.2, 1.5) {a)};
        \node at (4.6,0) {
            \includegraphics[scale=0.32]{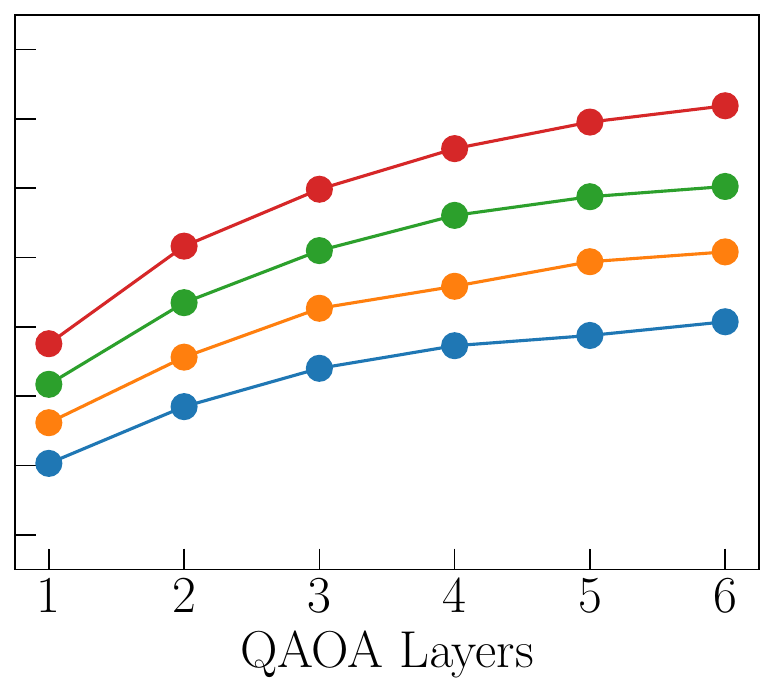}
        };
        \node at (3,1.5) {b)};
        \node at (0,-3.8) {
            \includegraphics[scale=0.32]{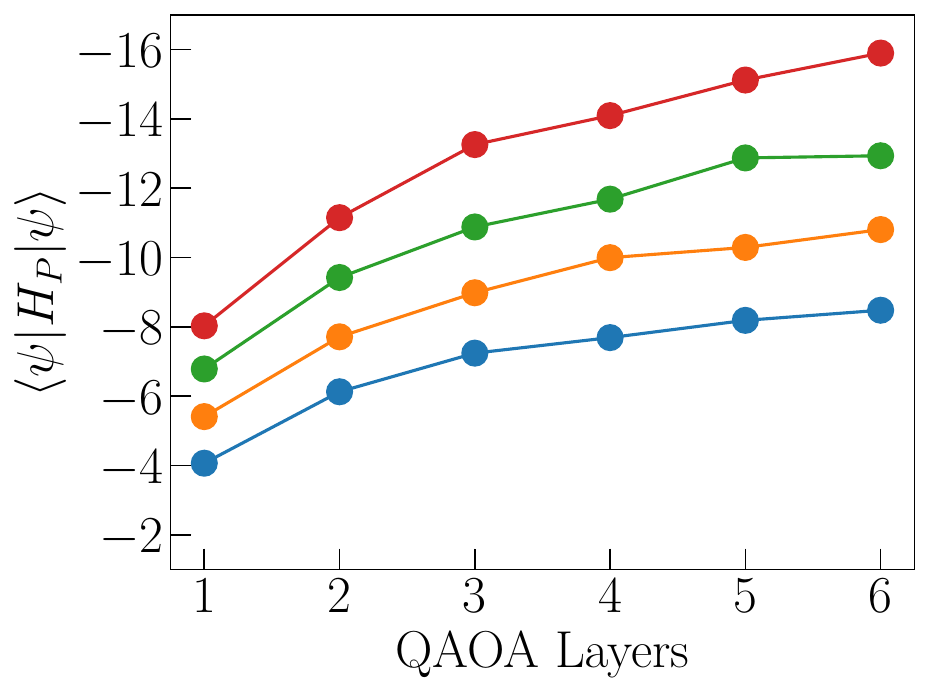}
        };
        \node at (-1.2,-2.3) {c)};
        \node at (4.5,-3.5) {
            \includegraphics[scale=0.38]{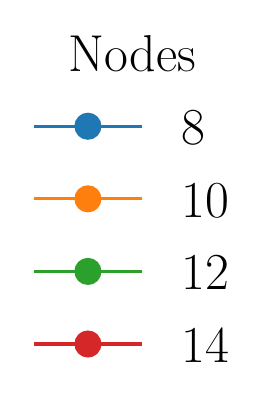}
        };
    \end{tikzpicture}
    \caption{The average cost over 100 graphs after the optimisation step of QAOA for a) complete, b) 3-regular and c) 4-regular graphs respectively for one to six QAOA layers. We find the cost after optimisation is always less than zero.}
    \label{fig:costs}
\end{figure}
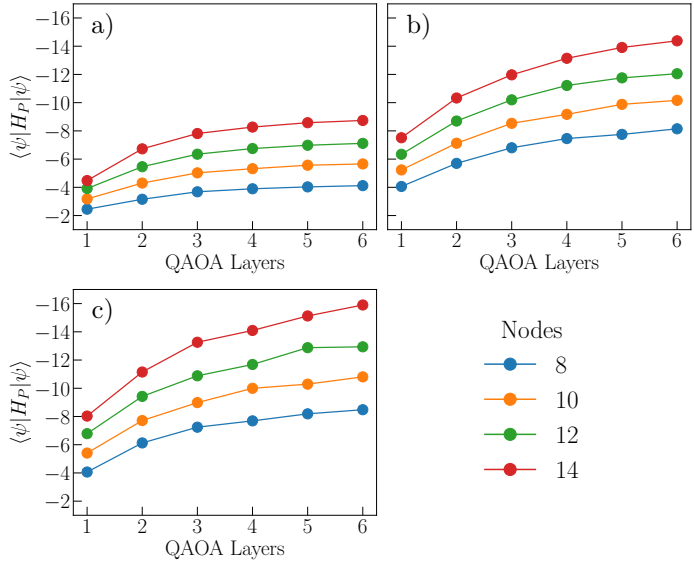
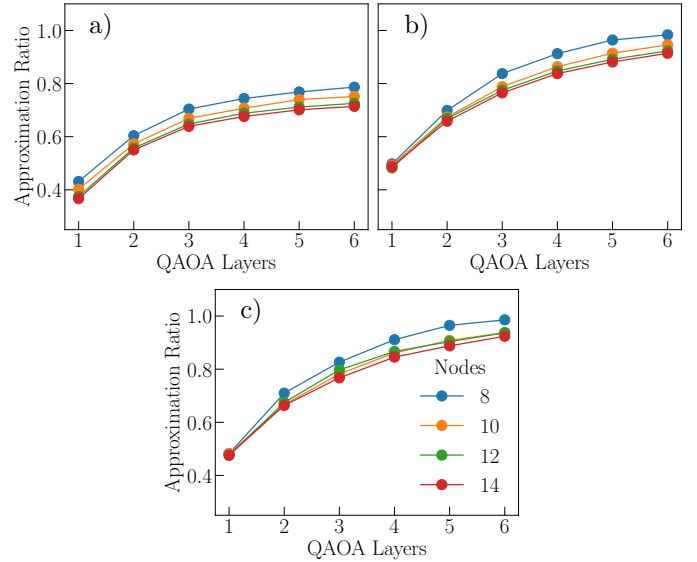
\begin{figure}
    \begin{tikzpicture}
        \node at (0,0) {
            \includegraphics[scale=0.32]{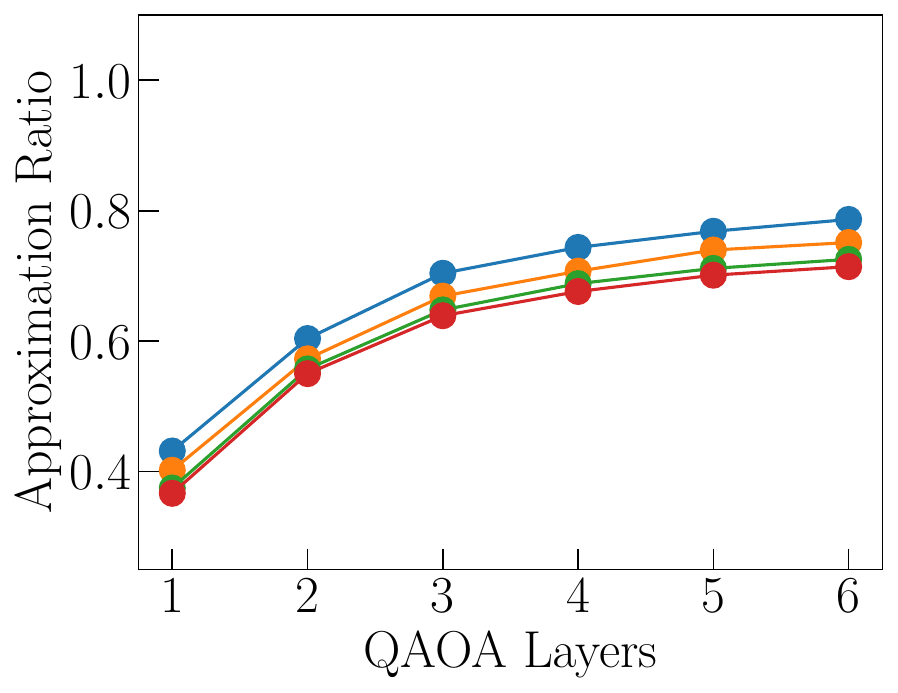}
        };
        \node at (-1.2,1.5) {a)};
        \node at (4.5,0) {
            \includegraphics[scale=0.32]{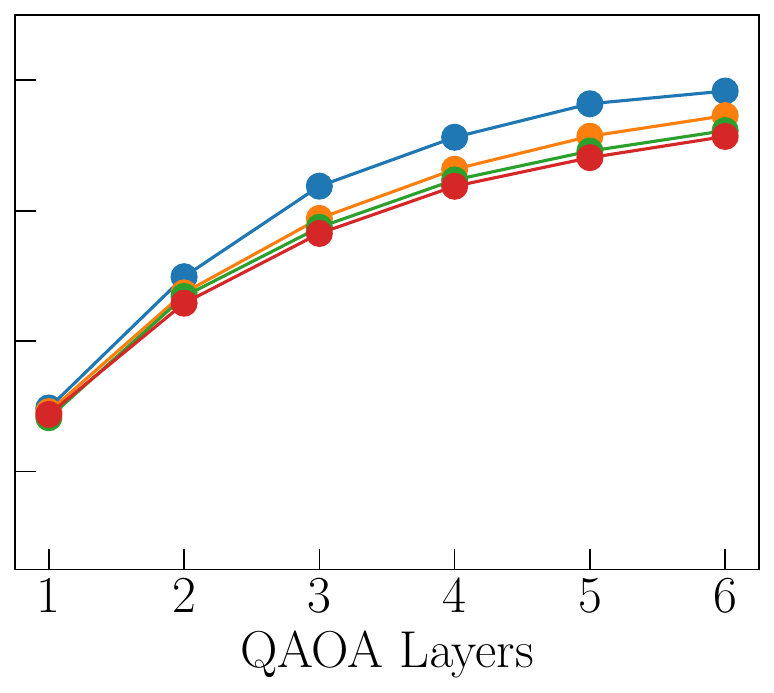}
        };
        \node at (3,1.5) {b)};
        \node at (2,-3.8) {
            \includegraphics[scale=0.32]{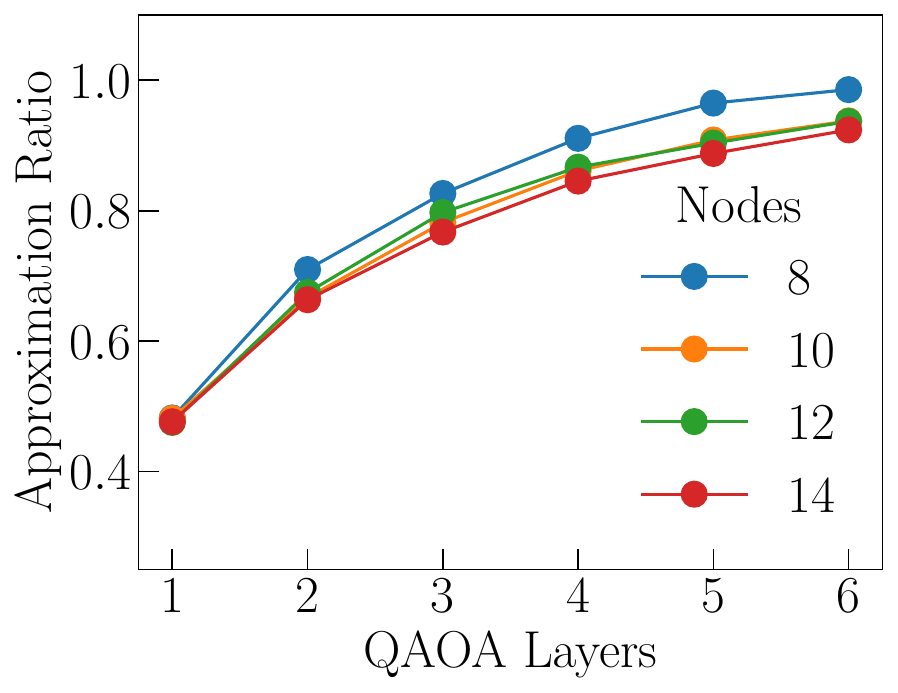}
        };
        \node at (0.8,-2.3) {c)};
    \end{tikzpicture}
    \caption{The approximation ratio for a) complete, b) 3-regular and c) 4-regular graphs. It is recognised that complete graphs with random weights appear to be harder for QAOA to solve with approximation ratios plateauing at $>0.8$. This is contrasted by the random regular graphs with unit edge weights which appear to have approximation ratios approaching one.}
    \label{fig:apr_reg}
\end{figure}
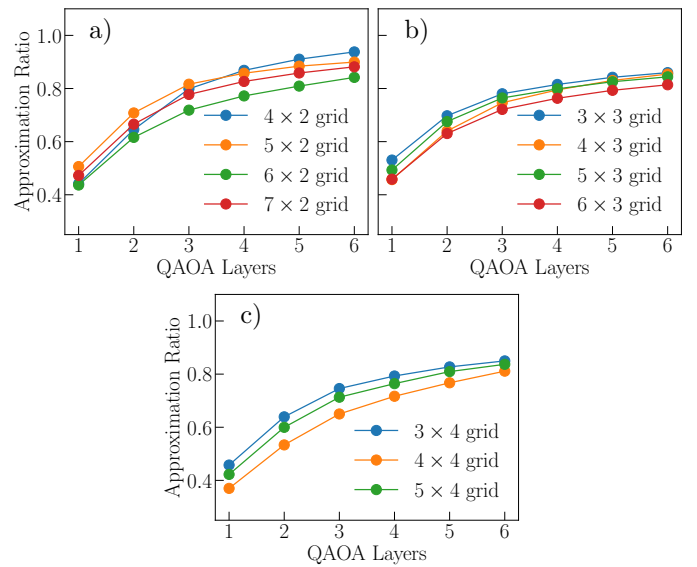
\begin{figure}
    \begin{tikzpicture}
        \node at (0,0) {
            \includegraphics[scale=0.32]{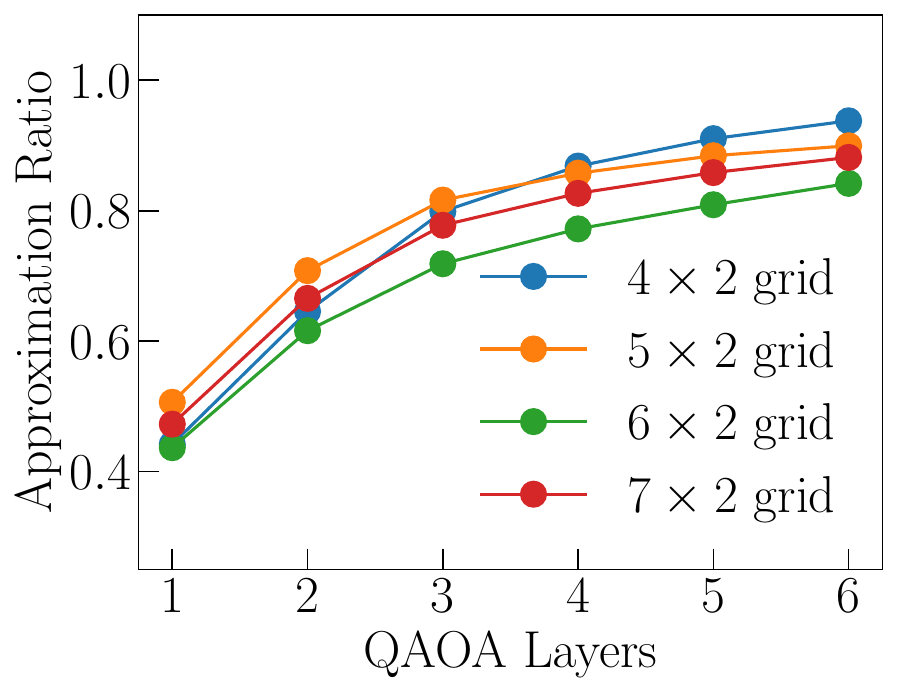}
        };
        \node at (-1.2, 1.5) {a)};
        \node at (4.5,0) {
            \includegraphics[scale=0.32]{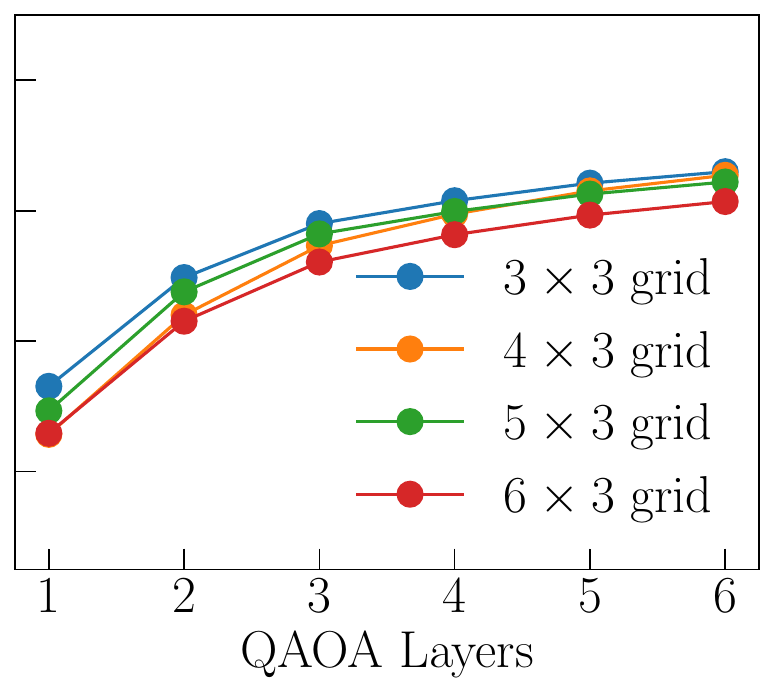}
        };
        \node at (3,1.5) {b)};
        \node at (2,-3.8) {
            \includegraphics[scale=0.32]{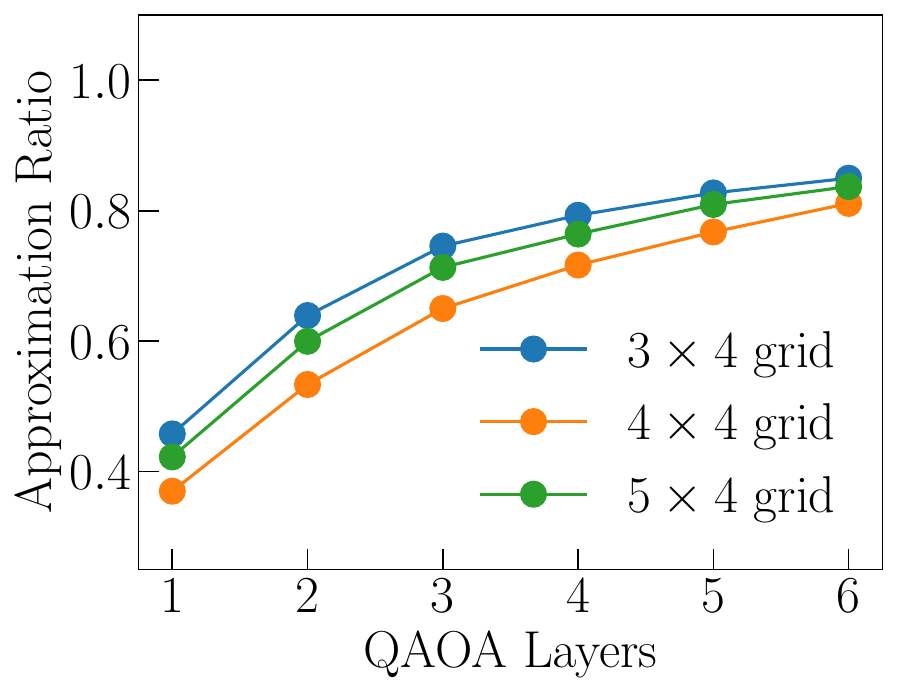}
        };
        \node at (0.8,-2.3) {c)};
    \end{tikzpicture}
    \caption{The approximation ratio for grids with a) two, b) three and c) four columns respectively for one to six QAOA layers. It is recognised that the approximation ratio for three and four-column grids appears to plateau at a lower value with two-column grids approaching an approximation ratio of 1, indicating that QAOA is able to determine the ground state of the Hamiltonian exactly.}
    \label{fig:apr_grids}
\end{figure}
\section{The QAOA Approximation Ratio} \label{apn_rat_def}
\begin{figure*}
    \begin{tikzpicture}
        \node at (0,0) {
            \includegraphics[scale=0.385]{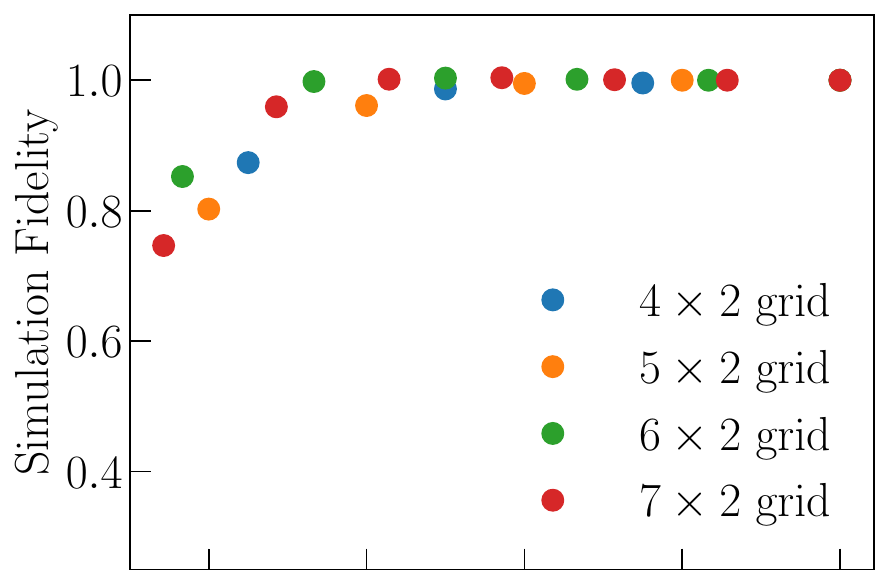}
        };
        \node at (5.5,0) {
            \includegraphics[scale=0.385]{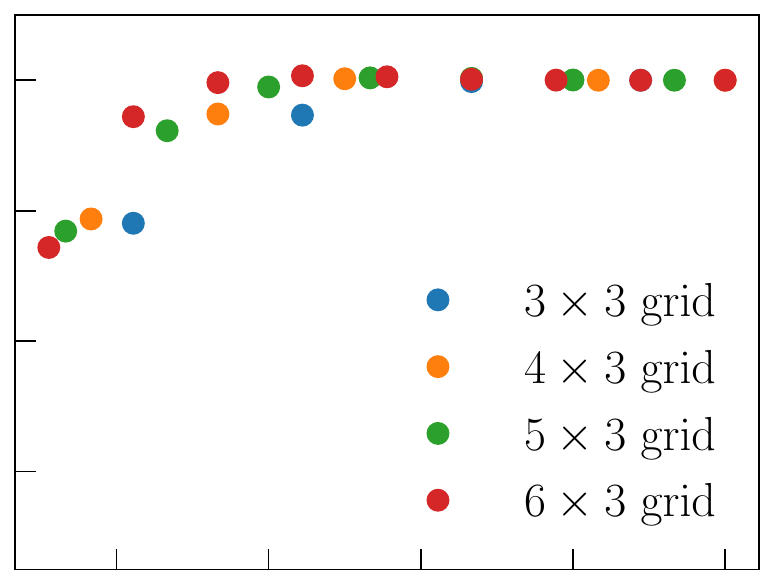}
        };
        \node at (11.15,-0.025) {
            \includegraphics[scale=0.385]{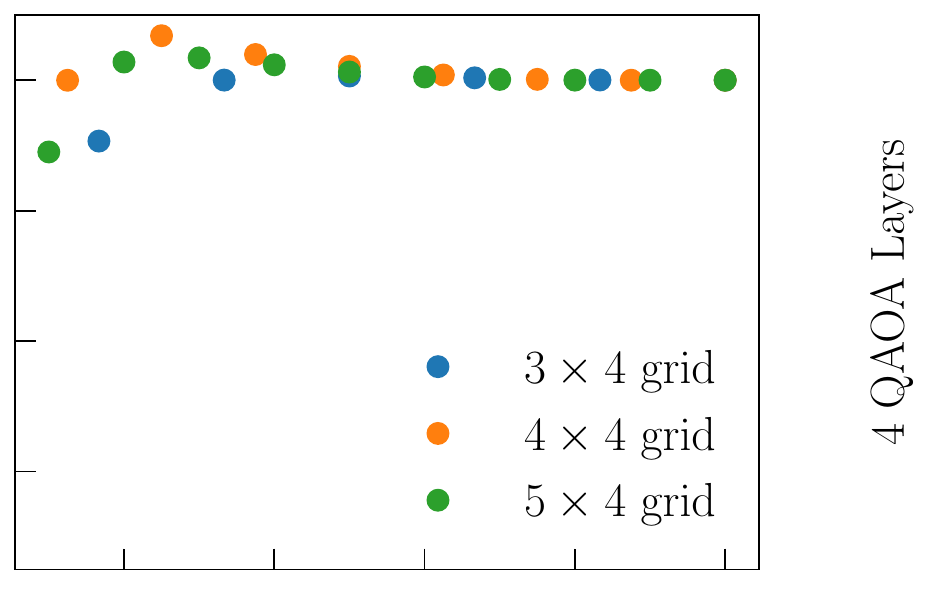}
        };
        \node at (0,-4.2) {
            \includegraphics[scale=0.385]{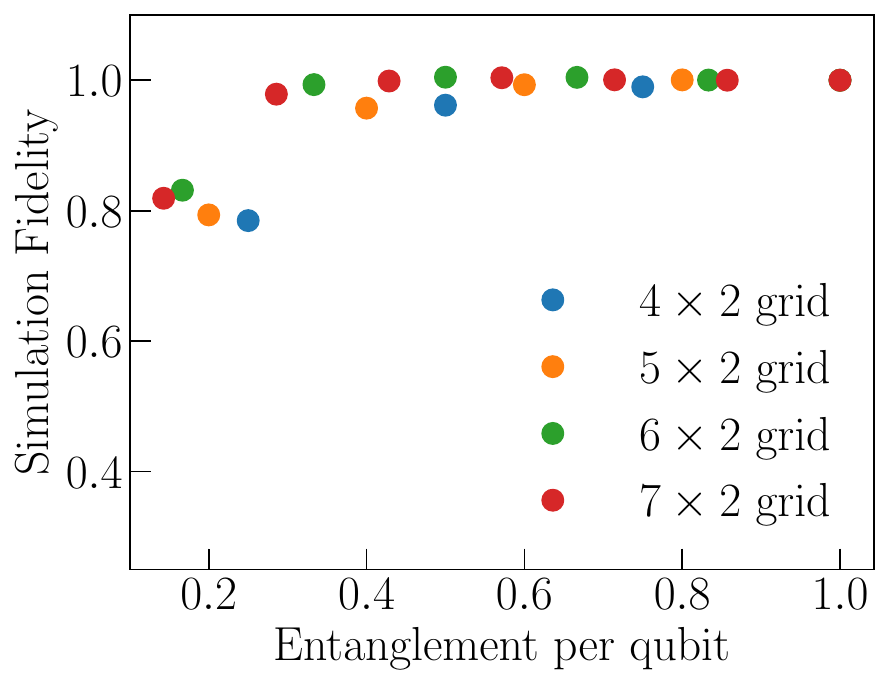}
        };
        \node at (5.5,-4.2) {
            \includegraphics[scale=0.385]{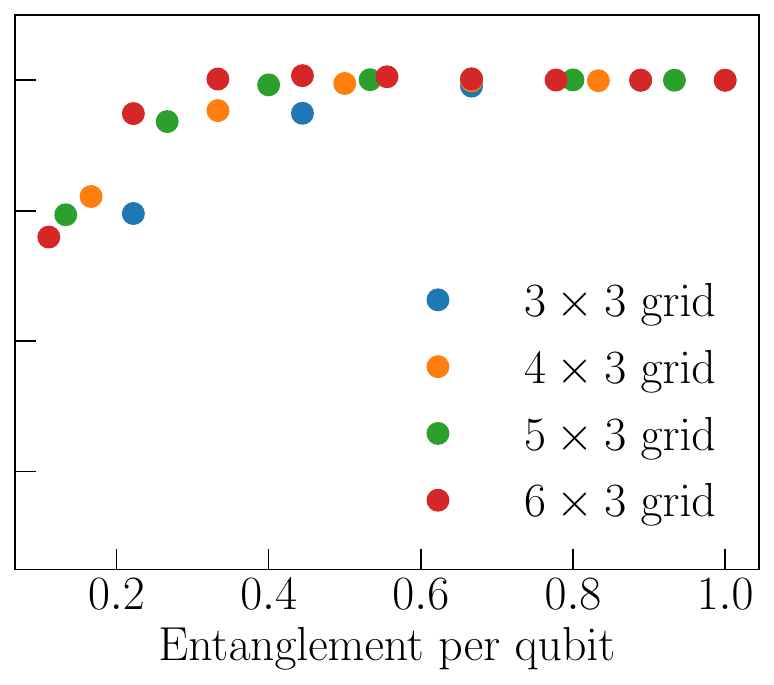}
        };
        \node at (11.15,-4.2) {
            \includegraphics[scale=0.385]{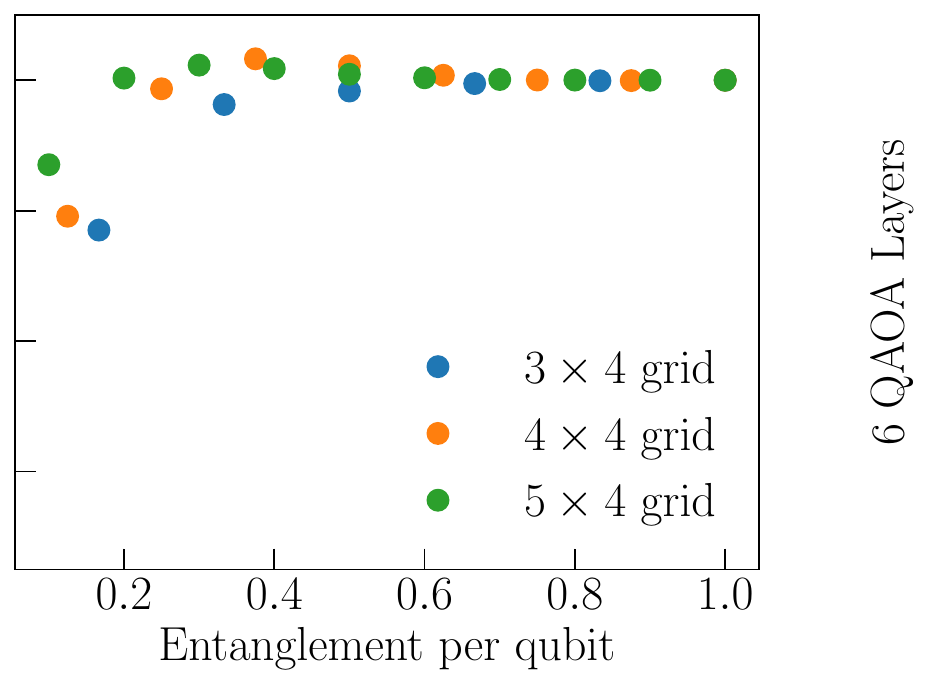}
        };
    \end{tikzpicture}
    \caption{The simulation fidelity at four QAOA layers (top row) and six QAOA layers (bottom row) for grid graphs with two (left column), three (middle column) and four columns (right column) respectively. We find across all graphs at both four and six layers that there is a strong resilience to truncation of entanglement. It is also recognised that similar to the QNN models trained on the CIFAR dataset, truncation sometimes results in an improvement in the cost compared to the untruncated state, resulting in a ``simulation fidelity'' of $>1$. Possible reasons for this are discussed in the main body of this work.}
    \label{fig:big_p_grids}
\end{figure*}
The performance of QAOA is commonly determined by the approximation ratio defined as\cite{farhi_quantum_2014},
\begin{equation}
    \text{Approximation Ratio} = \frac{\mel{\psi}{H_P}{\psi}}{\mel{\psi_{\text{Exact}}}{H_P}{\psi_{\text{Exact}}}}, \label{eq:apr_rat}
\end{equation}
where $\ket{\psi_{\text{Exact}}}$ is the exact ground state solution of $H_P$ defined in Section \ref{sec:qaoa_method} and $\ket{\psi}$ is the solution state found using QAOA. It is recognised that the approximation ratio as stated is ill-defined as (1) one may shift the energy spectrum by a constant which does not materially change the problem but will result in a different approximation ratio; (2) $\mel{\psi}{H_p}{\psi}$ can be any real number and as such the numerator and denominator may be of a different sign, resulting in a negative approximation ratio and; (3) $\mel{\psi_{\text{Exact}}}{H_P}{\psi_{\text{Exact}}}$ may be zero resulting in an undefined approximation ratio. A well-defined approximation ratio has been established in Ref. \cite{DEMANGE1996117} which introduces a third variable to Equation \eqref{eq:apr_rat} which corresponds to the worst case cost, $\mel{\psi_{\text{worst}}}{H_P}{\psi_{\text{worst}}}$. With this, the approximation ratio will not vary under a constant shift in the energy spectrum and will always be in $[0,1]$. It is recognised that for the MAX-CUT problem Hamiltonian $H_P=\frac{1}{2}\sum_{(i,j)\in E} w_{ij} (Z_iZ_j-I)$ the worst case scenario is $\mel{\psi_{\text{worst}}}{H_P}{\psi_{\text{worst}}}=0$ and as such one can continue to use the approximation ratio as defined in Equation \eqref{eq:apr_rat}. However, in previous works studying the entanglement of QAOA suing the MAX-CUT problem\cite{dupont_calibrating_2022,dupont_entanglement_2022} the Hamiltonian used is $H_P=\sum_{(i,j)\in E} w_{ij} Z_i Z_j$ where the worst case cost is $\abs{E}$. As such one must exercise caution when using the approximation ratio and MAX-CUT Hamiltonian as defined in this work. Recognising that $\mel{\psi_\text{Exact}}{H_P}{\psi_\text{Exact}} < 0$ for all graph instances considered in this work and that $\mel{\psi}{H_P}{\psi} < 0$, which we show in Appendix \ref{apn_sim_fid}, we do not alter the approximation ratio or Hamiltonian given in order to maintain comparability with previous works.

\section{QAOA Approximation Ratio Results} \label{apn_rat_res}
The approximation ratio is tracked for one to six QAOA layers for regular and complete graphs in FIG. \ref{fig:apr_reg} as well as grid graphs in FIG. \ref{fig:apr_grids}. These are the same graphs considered in FIG. \ref{fig:reg} and FIG. \ref{fig:grids} respectively in the main body of this work. We find that all graph types considered show a monotonic increase in the approximation ratio as the number of QAOA layers is increased, additionally noting that increasing the system size only has a negligible negative effect on the performance of QAOA.

Additionally, it is recognised that the approximation ratios for the complete graphs in FIG. \ref{fig:apr_reg}(a) appear to be significantly lower than those of random regular graphs of the same size in FIG. \ref{fig:apr_reg}(b-c) with the complete graphs plateauing at $0.8$ compared to close to one for the regular graphs. This is generally an indication that the complete graphs are more difficult to solve using QAOA compared to the less dense regular graphs. This is also consistent with the results in the main body of this work which found that the entanglement entropy of the complete graphs continued to increase up to six QAOA layers, compared to regular graphs which had a maximum entanglement entropy at two or three layers.

Regarding the grid graphs, it is found that the approximation ratio for two-column grids in FIG. \ref{fig:apr_grids}(a) are similar to the random regular graphs, whereas three and four-column grids appear to have a lower approximation ratio compared to random 4-regular graphs as well as the two-column grids. This is consistent with the main body of this work which finds that the entanglement entropy of the three and four-column grid graphs is higher compared to two-column grids. However, we do note that the entanglement entropy of the random 4-regular graphs is similar to that of the three and four-column grids despite the difference in approximation ratio. 

Overall the approximation ratio serves as an adequate measure in determining the performance of QAOA provided that one is careful with the formulation of the problem Hamiltonian. In that regard, in its most commonly used form in the study of QAOA and VQAs in general, one needs to be careful when using it to perform fundamental analysis as it is ill-defined if considering various problem Hamiltonians. For the types of problem considered in this work, the approximation ratio primarily served as a tool in determining whether QAOA is performing well or not, finding that the performance of QAOA as determined by the approximation ratio appears to be loosely tied to the entropy of entanglement of the final state.
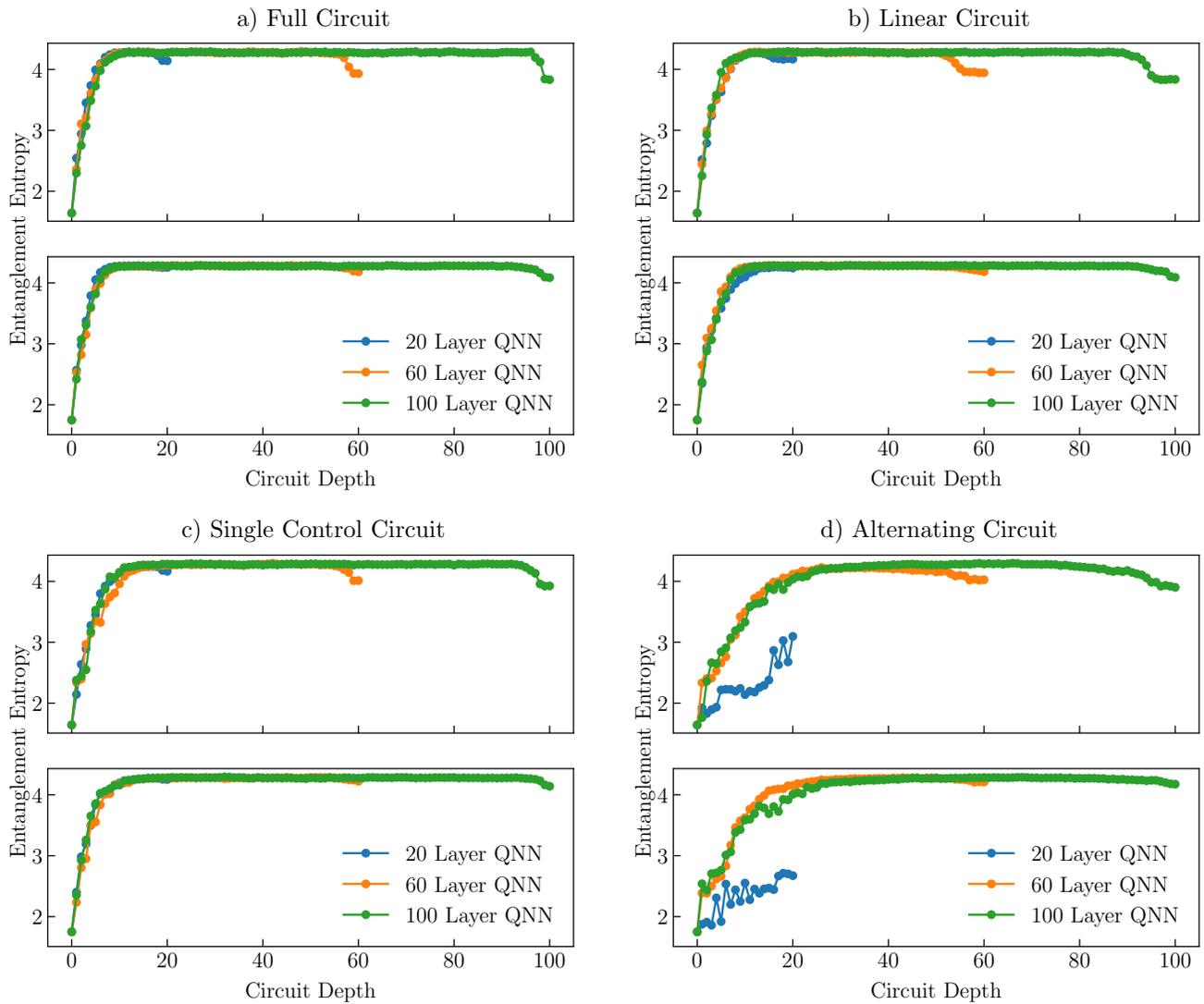
\begin{figure*}
    \begin{tikzpicture}
      \node at (0,0) {
                \includegraphics[scale=0.6]{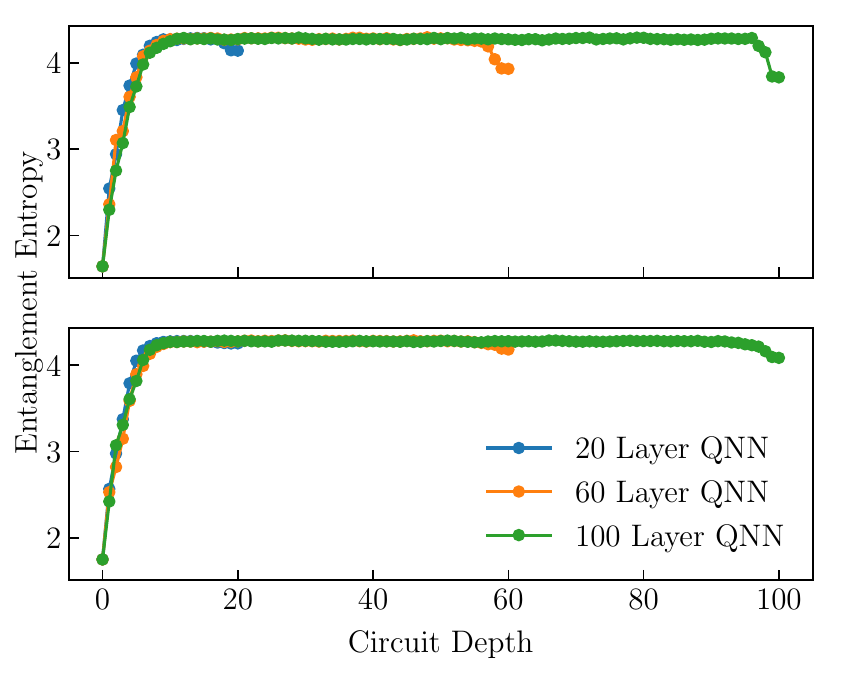}
            };       
      \node at (0.2,3.5) {a) Full Circuit};
    \end{tikzpicture}
    \begin{tikzpicture}
      \node at (0,0) {
                \includegraphics[scale=0.6]{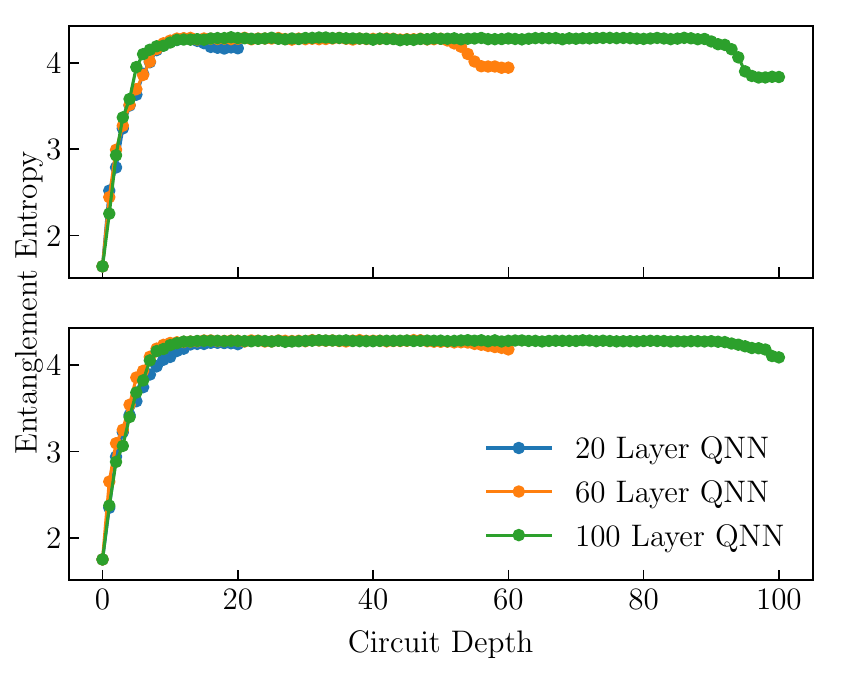}
            };       
      \node at (0.2,3.5) {b) Linear Circuit};
    \end{tikzpicture}
    \begin{tikzpicture}
      \node at (0,0) {
                \includegraphics[scale=0.6]{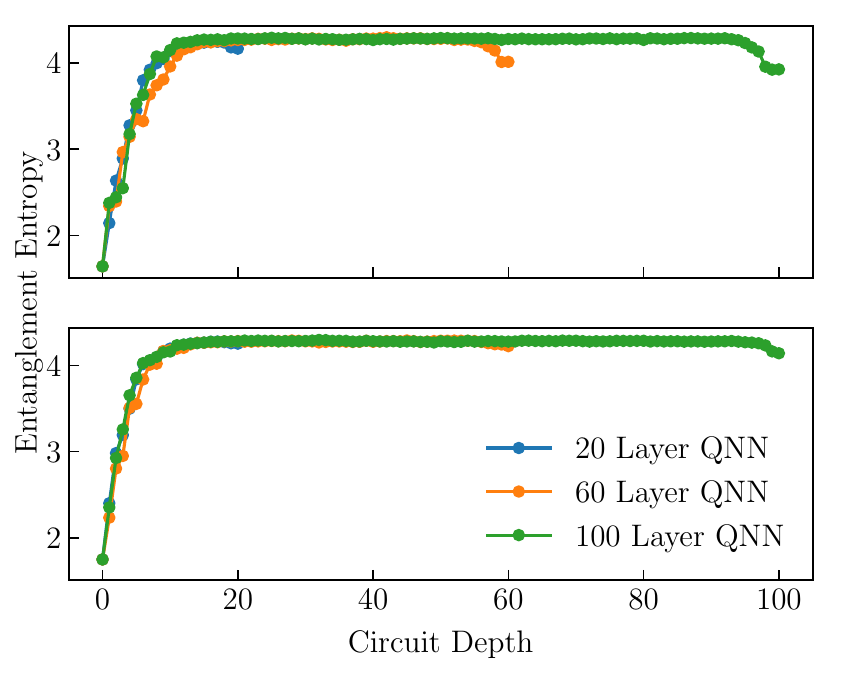}
            };       
      \node at (0.2,3.5) {c) Single Control Circuit};
    \end{tikzpicture}
    \begin{tikzpicture}
      \node at (0,0) {
                \includegraphics[scale=0.6]{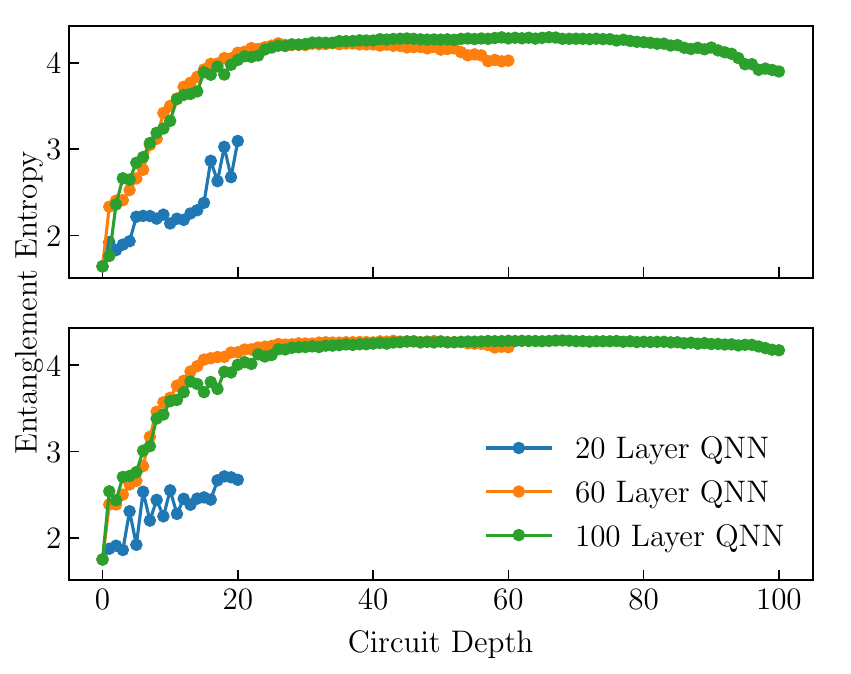}
            };       
      \node at (0.2,3.5) {d) Alternating Circuit};
    \end{tikzpicture}
    \caption{The evolution of entanglement throughout the QNN circuit for the remaining ansätze not shown in the main body of this work. The top sub-figures are for circuits trained on the BMNIST dataset with the bottom sub-figures corresponding to circuits trained on the MNIST dataset. We observe that there is a minor plateau towards the end of the BMNIST circuits for all ansätze shown, which was not observed in the periodic ansatz. As the $20$ layer alternating ansatz circuit is not capable of generating maximal entanglement, the entanglement entropy throughout the circuit appears to evolve differently compared to the other ansätze.}
    \label{fig:alls}
\end{figure*}
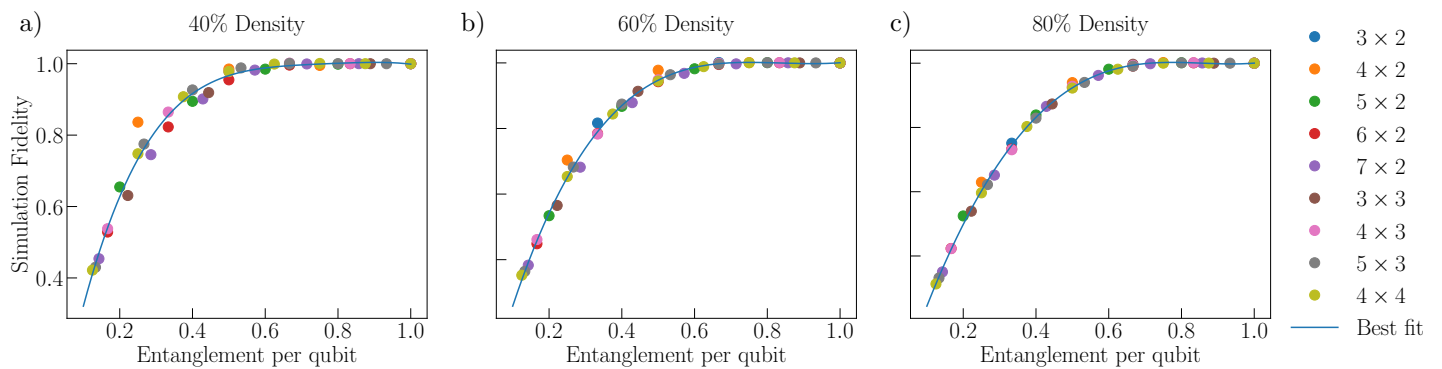
\begin{figure*}
    \begin{tikzpicture}
        \node at (0,0) {
            \includegraphics[scale=0.38]{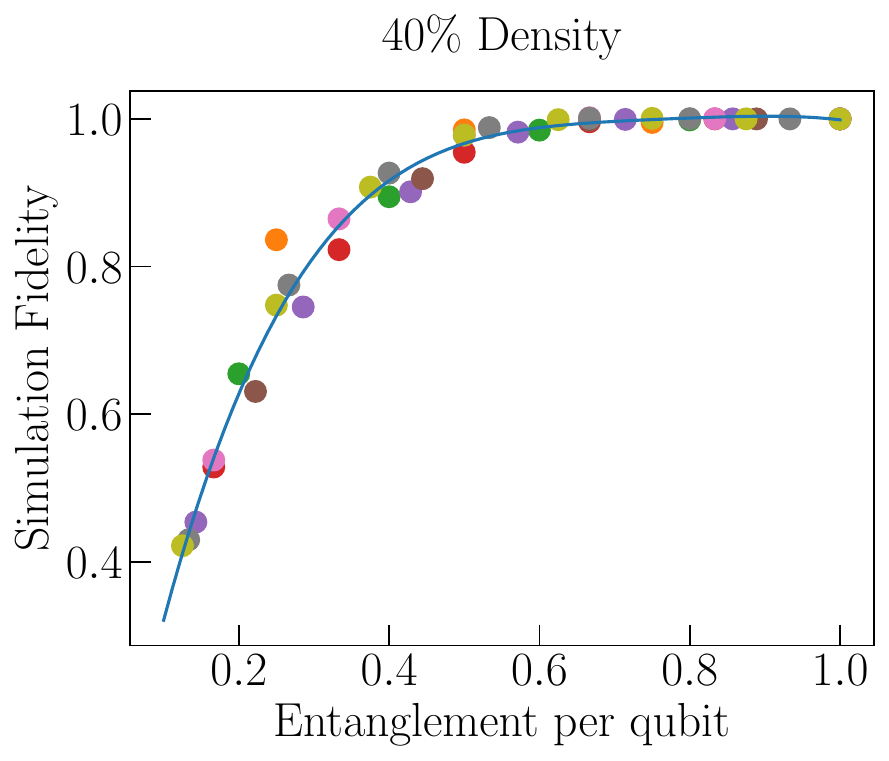}
        };
        \node at (-2.5,2.2) {a)};
    \end{tikzpicture}
    \begin{tikzpicture}
        \node at (0,0) {
            \includegraphics[scale=0.38]{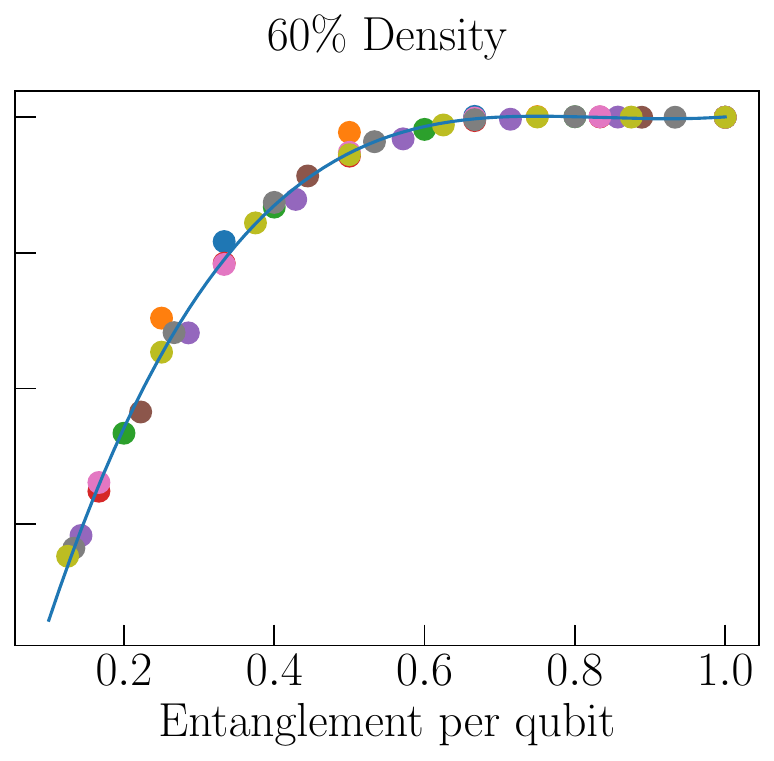}
        };
        \node at (-2.7,2.2) {b)};
    \end{tikzpicture}
    \begin{tikzpicture}
        \node at (0,0) {
            \includegraphics[scale=0.38]{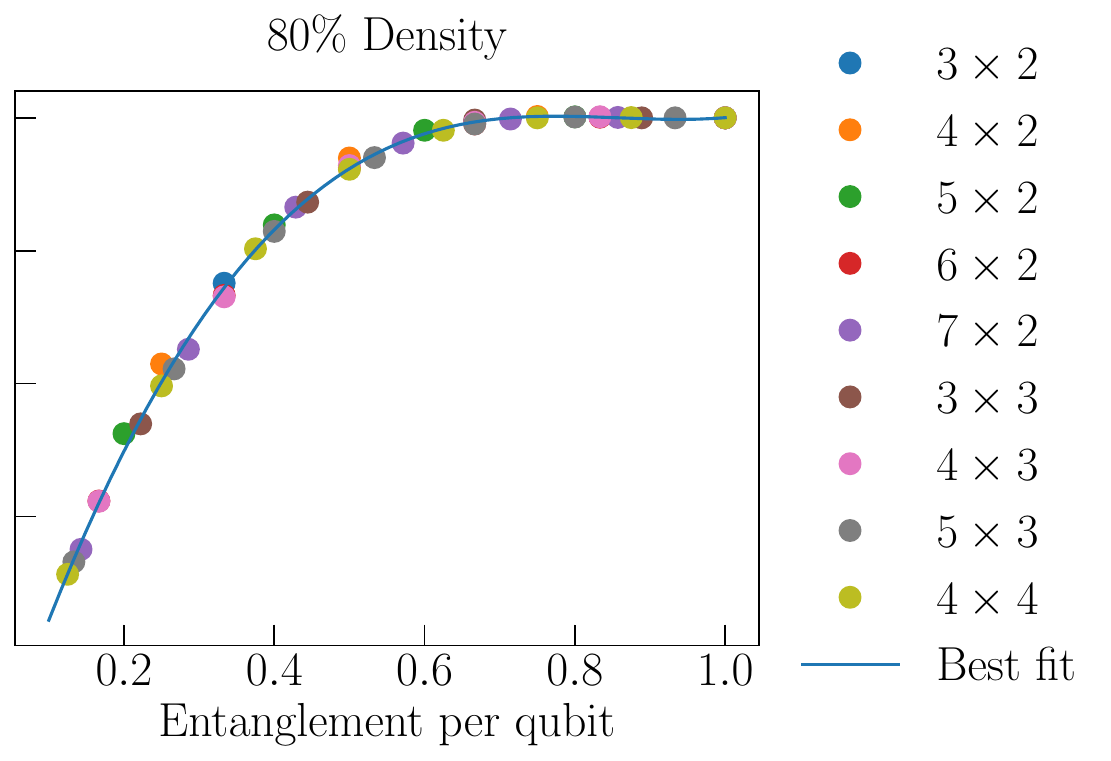}
        };
        \node at (-3.6,2.2) {c)};
    \end{tikzpicture}
    \caption{The simulation fidelity of graphs with density a) $40\%$, b) $60\%$ and c) $80\%$ at two QAOA layers including a polynomial fit of degree 4 as also shown in FIG. \ref{fig:trans} of the main text. Note that the scaling of the fidelity becomes more regular as the density of the graph increases and the edges become less structured.}
    \label{fig:trans_raw}
\end{figure*}
\section{Simulation Fidelity for larger QAOA Simulations on Grid Graphs} \label{apn_big_qaoa}
We provide the simulation fidelity for the grid graphs discussed in the main body of this work at four and six QAOA layers in FIG. \ref{fig:big_p_grids}. It is noted that the scaling appears to be similar to that of the 3 and 4-regular graphs at the same number of QAOA layers. The primary distinction is that there appears to be a notable increase in the ``simulation fidelity'' upon truncation. Similar to the QNNs trained on the CIFAR dataset, it is not apparent why this occurs, but it is likely due to the act of truncation resulting in the state accessing a part of the Hilbert Space that it previously could not with the unitary operations which QAOA uses.
\begin{figure}
    \begin{tikzpicture}
        \node at (0,0) {
            \includegraphics[scale=0.32]{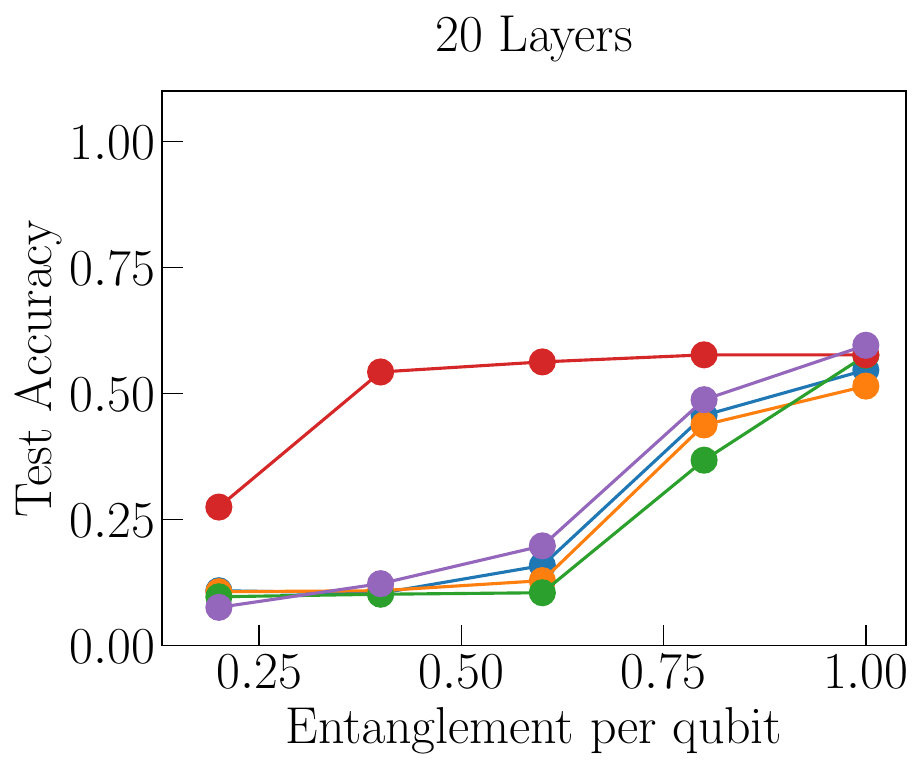}
        };
        \node at (4.55,0) {
            \includegraphics[scale=0.32]{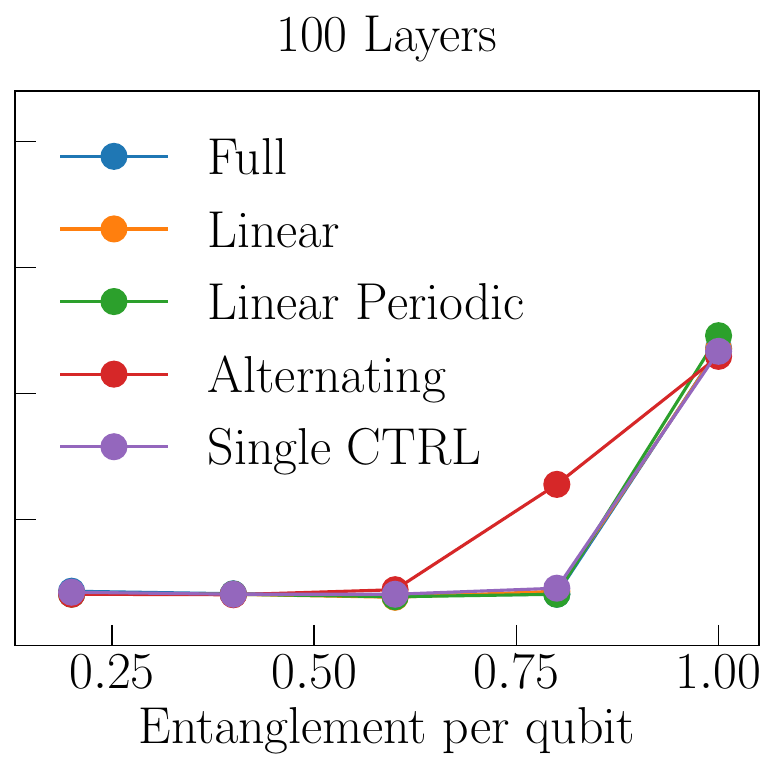}
        };
    \end{tikzpicture}
    \caption{The test accuracy at $20$ and $100$ layers for the $28 \times 28$ greyscale FMNIST dataset. The results for this dataset appear to closely resemble those of the MNIST dataset. The results at $60$ layers are shown in FIG. \ref{fig:s_test_accs} along with the other datasets considered in this work.}
    \label{fig:s_test_accs_fmnist}
\end{figure}
\begin{figure}
    \begin{tikzpicture}
        \node at (0,0) {
            \includegraphics[scale=0.32]{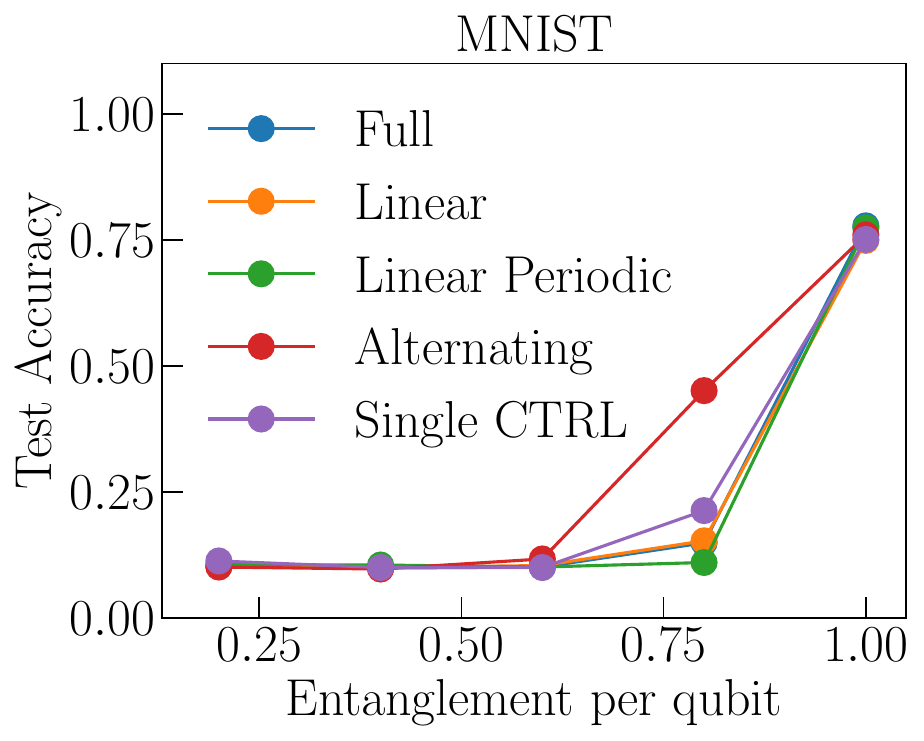}
        };
        \node at (4.55,0) {
            \includegraphics[scale=0.32]{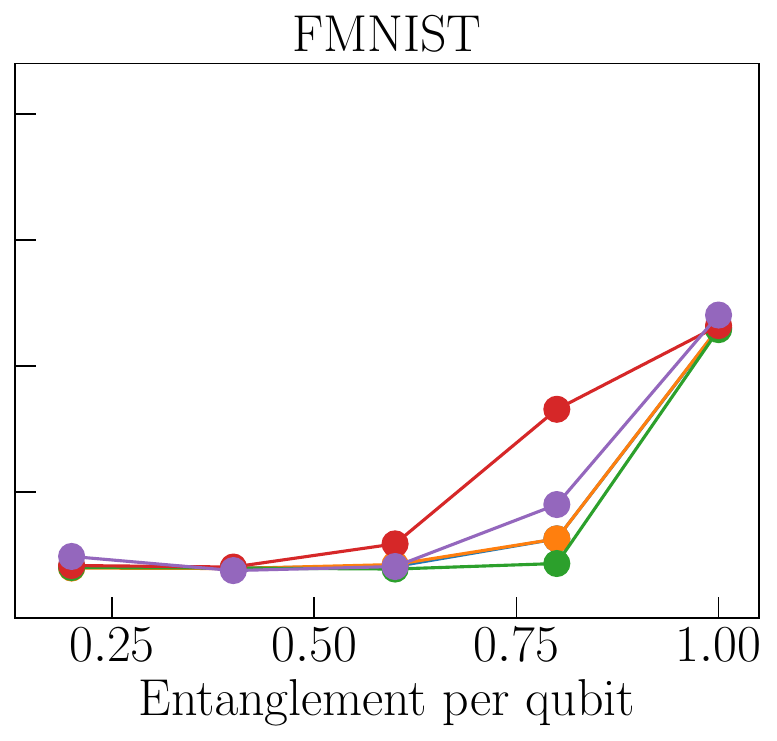}
        };
        \node at (0,-4) {
            \includegraphics[scale=0.32]{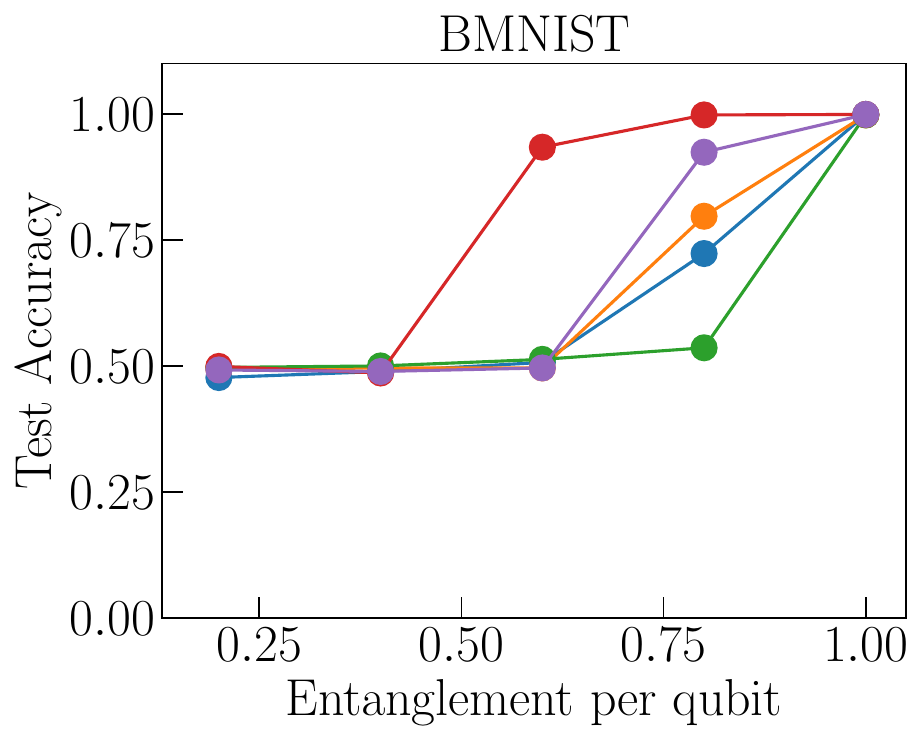}
        };
        \node at (4.55,-4) {
            \includegraphics[scale=0.32]{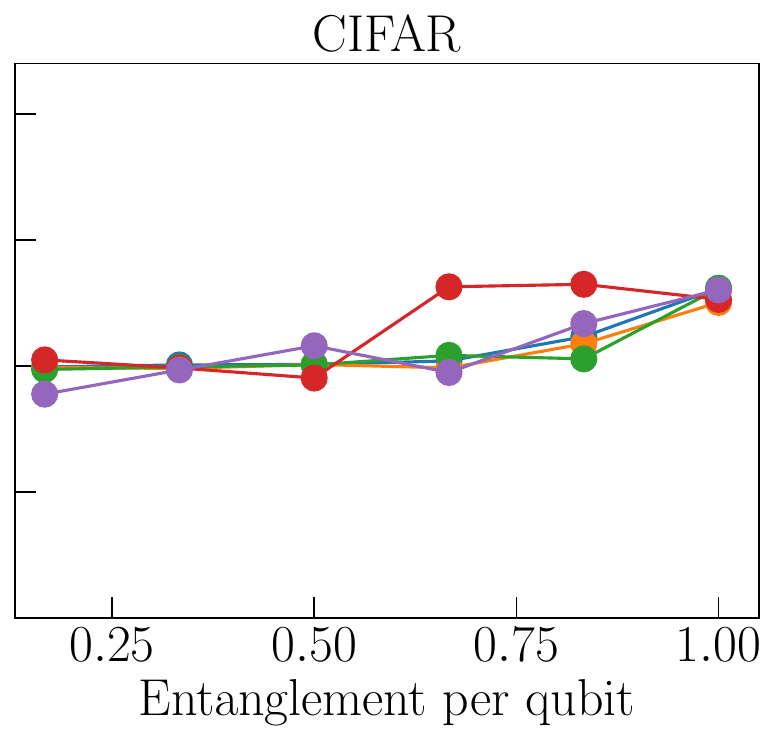}
        };
    \end{tikzpicture}
    \caption{The test accuracy for 60-layer QNNs for all datasets considered in this work. Compared to FIG. \ref{fig:test_accs} there appears to be a general decrease in the resilience towards truncation compared to the 20-layer QNNs across all datasets. Like the 100-layer QNN for the CIFAR dataset there appears to be a small increase in the test accuracy when a small amount of truncation is performed.}
    \label{fig:s_test_accs}
\end{figure}
\section{A note on the alternating ansatz} \label{apn_alt}
Consider the fundamental building blocks of the alternating ansatz, firstly,
\begin{equation}
    \begin{quantikz}[row sep={2mm}, column sep=0.1cm]
        \ctrl{1} & \qw{} \\
        \targ{} & \targ{} \\
        \qw{} & \ctrl{-1}
    \end{quantikz}
    =
    (CNOT \otimes I)(I \otimes NOTC).
\end{equation}
It is apparent that the CNOT gates in this configuration commute as the Pauli $X$ operation that may be conditionally applied to the second qubit commutes with itself. Likewise for the second building block,
\begin{equation}
    \begin{quantikz}[row sep={2mm}, column sep=0.1cm]
        \targ{} & \qw{} \\
        \ctrl{-1} & \ctrl{1} \\
        \qw{} & \targ{}
    \end{quantikz}
    =
    (NOTC \otimes I)(I \otimes CNOT),
\end{equation}
where it is recognised that the control operation likewise commutes with itself. As such it becomes apparent that all the CNOT gates in the alternating ansatz commute with each other. Given this, one can conclude that the state resulting from this ansatz lives in a greatly restricted Hilbert space.  This may be compared to the other ansätze considered in this work where the CNOT operations do not commute with each other to such an extent. An additional consequence of this is that the entanglement that this ansatz can produce is restricted at low depth as the only source of non-commutativity are the single qubit rotations which sandwich the CNOT gates.
\section{Tracking Entanglement Entropy for all Ansätze} \label{apn_qnn_entrop}
For completeness, we include the evolution of entanglement entropy throughout the circuit for the QNNs trained on the MNIST and BMNIST which are not included in the main body of this work in FIG. \ref{fig:alls}. There does not appear to be a substantial difference between the evolution of the periodic ansatz and that of the full, linear and single control ansätze. Regarding the alternating ansatz, the only difference appears to be with the $20$ layer QNN where it is recognised that the ansatz cannot generate very high entanglement and as such QNN is in a different regime compared to other ansätze which can, or the alternating ansatz at higher depth.

\section{Additional data for graphs of varying edge density} \label{apn_raw_den}
FIG. \ref{fig:trans_raw} shows the average simulation fidelity at each entanglement per qubit for graphs of approximately $40\%$, $60\%$ and $80\%$ density, analogous to that shown in FIG. \ref{fig:trans} of the main text. It is observed that as the density increases there appears to be less variation in the scaling of the simulation fidelity. This is similar to the results seen for complete graphs as per FIG. \ref{fig:reg}.

\section{Additional QNN Test Accuracies} \label{apn_qnn}
For completeness, we include the test accuracy at each entanglement per qubit for the $28 \times 28$ greyscale FMNIST dataset in FIG. \ref{fig:s_test_accs_fmnist}. The results for this dataset mimic the similar MNIST dataset where it is found that resilience towards truncation reduces significantly for larger QNN models with only a modest increase in test accuracy.

Additionally, we have included the test accuracy for each entanglement per qubit at 60 layers in FIG. \ref{fig:s_test_accs}. The results continue the trend seen in the main body of this work where resilience towards truncation continues to decrease across all datasets as the number of layers in the model increases.
\end{document}